\documentclass[useAMS,usenatbib]{mn2e}
\usepackage{graphicx}

\title[Mergers and Feedback]{A Panchromatic Survey of Post-starburst Mergers: searching
for feedback}
\author[R. De Propris  and J. Melnick]{Roberto De Propris$^{1}$\thanks{E-mail:
rodepr@utu.fi} and Jorge Melnick$^{2}$\\
$^{1}$ Finnish Centre for Astronomy with ESO, University of Turku, V{\"a}is{\"a}l{\"a}ntie,
Piikki{\"o}, 21500, Finland\\
$^{2}$ European Southern Observatory, Av. Alonso de Cordova 3107, Santiago 19, Chile}
\begin{document}

\date{}

\pagerange{\pageref{firstpage}--\pageref{lastpage}} \pubyear{2002}

\maketitle

\label{firstpage}

\begin{abstract}

We consider the morphology, stellar populations, structure and AGN activity of 10 post-starburst
(K+A) galaxies with HST observations, full spectral coverage in the optical, spectral energy 
distributions from 0.2 to 160 $\mu$m, X-ray and radio data. Our results show that the PSG
phenomenon is related to mergers and interactions, and that star formation was likely triggered
during close passes prior to final coalescence. We performed a detailed qualitative analysis of the 
observed light distribution, including low-surface brightness tidal features and color profiles, in 
high-resolution multi-band imaging with HST. We find evidence that star formation was
centrally concentrated and that quenching took place from the inside-out, consistent with
the occurrence of a feedback episode. Most of our PSGs contain massive bulges and therefore
should host supermassive black holes. We search for AGN activity in spectra (line ratios),
optical variability, X-ray emission at 0.5--7.0 KeV and radio emission at 20cm: all four lines
of evidence show there is no active AGN accreting at more than 0.1\% of the Eddington
luminosity. We conclude that mergers may be a necessary, but not a sufficient condition, for
AGN activity and that they are not likely to be important in our objects. If PSGs are good test 
cases for quenching and evolution to the red sequence, AGNs may play a smaller role than
expected.

\end{abstract}


\section{Introduction}
Post-starburst galaxies (hereafter PSGs, sometimes called E+A or K+A galaxies from their distinctive 
spectral signatures) are observed shortly ($\sim 0.5$ Gyr) after a significant episode of star formation but 
show no evidence that they are currently forming new stars \citep{Du2010,Shin2011,Nielsen2012,
Melnick2013}. These galaxies are the best local examples of systems observed crossing the 'green valley'
between the two main groupings of galaxies in colour space (the so-called blue cloud and red sequence) 
and may allow us to understand how star formation is started and terminated in galaxies and thus provide 
a clue to the origin of the persistent bimodality in galaxy colours \citep{Strateva2001,Baldry2004,Wetzel2012}.

Simulations indicate that PSGs may be produced in mergers between gas-rich progenitors 
\citep{Bekki2005,Wild2009,Snyder2011}. Observations support the view that mergers and interactions 
have played an important role in these galaxies: \cite{Zabludoff1996} found that 5 of the 21 PSGs
in the Las Campanas Redshift Survey were morphologically disturbed, while investigation of these objects
with the Hubble Space Telescope (HST) by \cite{Yang2008} showed that all were interacting or
had undergone mergers. \cite{Goto2005} argued that 1/3 of all PSGs in the Sloan Digital Sky
Survey presented tidal tails or bright compact nuclei, indicative of current and recent mergers or
interactions, while \cite{Yamauchi2008} calculated that PSGs are half again as likely to have a
companion than a comparison sample of more normal (spectroscopically) galaxies. Although 
other mechanisms, such as ram stripping, are possible, especially in cluster environments \citep{
Ma2008, Yoshida2008, Wu2013}, \cite{Nolan2007} conclude that a merger/interaction channel 
must be the dominant pathway involved in the formation of PSGs (as do \citealt{Wu2013} in their
analysis of PSGs in a $z=0.9$ supercluster).

Mergers and interactions are expected to drive gas and dust towards the nuclear regions of galaxies,
where collisions and shocks induce star formation \citep{Hernquist1989,Joseph1995,Barnes1996,
Hopkins2009}. Many of the PSGs observed \cite{Yang2008} are seen to have positive colour gradients 
(i.e., bluer inward) or blue cores, unlike the mild negative gradients observed in normal galaxies
\citep{LaBarbera2010}. \cite{Pracy2005,Pracy2012,Pracy2013} and \cite{Goto2008a} also found 
gradients in H$_{\delta}$ absorption in integral field spectroscopy of PSGs, indicative of centrally 
concentrared star formation, supporting a scenario where star formation took place mostly in galaxy 
cores and these objects are now observed in an early post-merger stage.

We also expect that mergers contribute to the growth of bulges (e.g., \citealt{Toomre1977,
Zavala2012} and references therein) and that bulge growth signals or induces quenching
of star formation \citep{McGee2011,Bell2012}. In agreement with this, many PSGs have 
important bulge components \citep{Quintero2004,Yang2008}. We therefore expect that
PSGs also host supermassive black holes in their cores \citep{Antonucci1993, Kormendy1995}. 
This may provide a mechanism for the rapid cessation of star formation in these galaxies
\citep{Kaviraj2007,Choi2009,Melnick2013} via 'feedback' from the onset of
AGN activity \citep{Springel2005}. In the scenario for co-evolution of black holes and
galaxies described by \cite{Hopkins2008a,Hopkins2008b}, each galaxy undergoes a
series of mergers, which induce starbursts that are subsequently quenched by AGNs,
with the remnant settling on the red sequence. PSGs closely resemble some of the
intermediate evolutionary phases in these models, where AGN activity reaches its peak at 
or closely after the epoch of close approach or final coalescence. 

However, \cite{Shin2011} and \cite{Nielsen2012} find little evidence of 20cm emission in
PSGs from the \cite{Goto2007} sample at levels inconsistent with the presence of star formation
or AGNs. \cite{Liu2007} discovered evidence of a weak AGN in a nearby PSG. \cite{Georgakakis2008}
argued for the presence of AGNs in at least some $z \sim 1$ PSGs, while \cite{Brown2009}
find that they are only important for rather massive systems ($M_R > -22$). \cite{Wild2009} do
not detect evidence that AGN feedback has been important in their sample of $z \sim 0.8$
PSGs, while \cite{Wong2012} find only 2 LINERS (not necessarily AGNs) in their sample of
80 local PSGs. In our previous paper \citep{Melnick2013} we used line ratios from SDSS 
spectra for 808 PSGs and could not find convincing evidence for even weak AGNs. On the other hand,
\cite{Tremonti2007} measured powerful outflows from 10 of their 14 $z \sim 0.6$ PSGs, but
\cite{Coil2011} showed that these do not need to be powered by AGNs.

Here we study a sample of 10 nearby K+As observed in several stages of the merger process. 
We use spectra and photometry to reconstruct their star formation history. We examine deep 
Hubble Space Telescope (HST) images of these galaxies to identify tidal features and set
limits to their merging activity.  We use these deep, high-resolution images to study the distribution
of stellar populations, placing our galaxies on an approximate sequence joining merger stage and
star formation history. We then search for QSO activity in our objects from the optical, radio and
X-rays.

The following section describes the dataset and its properties. We comment on morphologies and colour 
gradients in section 3. The stellar populations are analysed in section 4 and the X-ray properties in section 5. 
We discuss these results in the light of the above models for AGN feedback in section 6. We adopt the conventional 
cosmology from the latest results of the WMAP 9 year analysis \citep{Hinshaw2013}.

\section{Dataset and Analysis}

We have selected galaxies from our original sample of PSGs in \cite{Melnick2013}, which is largely
based on the PSGs selected by \cite{Goto2007}. These are chosen from the SDSS DR7 with the 
conventional definition of a PSG, with equivalent widths of $H_{\alpha} > -3.0$ \AA, $H_{\delta} > 5.0$
\AA, and $[OII] > -2.5$ \AA\ (emission lines are negative). Apart from SDSS spectroscopy and optical
photometry \citep{York2000,Abazajian2009}, all these galaxies had FUV (1300 \AA) and NUV (2300 \AA)
photometry from the All-Sky, Medium and Deep Imaging Surveys on the GALEX telescope \citep{
Morrissey2007} as well as infrared ($YJHK$) data from either the 2 Micron All Sky Survey \citep{
Skrutskie2006} and the latest data release of the UKIDSS survey\footnote{The UKIDSS project is defined
in Lawrence et al (2007). UKIDSS uses the UKIRT Wide Field Camera (WFCAM; Casali et al. 2007). 
The photometric system is described in Hewett et al. (2006), and the calibration is described in Hodgkin 
et al. (2009). The pipeline processing and science archive are described in Irwin et al (2009, in prep) and
Hambly et al (2008)}, which explicitly targets the DR7 footprint. Mid-infrared imaging at 3.4, 4.5, 12.0
and 22.0 $\mu$m is provided by the WISE survey \citep{Wright2010}. All objects are also within the Faint
Images of the Radio Sky at 20cm survey (FIRST -- \citealt{Becker1995}).

One of our aims here is to use high quality (possibly multicolour) imaging to study the morphologies of
PSGs, identify merger signatures and analyse the distribution of the younger stellar populations. We 
searched the HST archive and found 8 galaxies with imaging in at least two colours ($B_{F438W}$
and $r_{F625W}$, with one having also $U_{F336W}$) from Program 11643 (PI: Zabludoff). 

We also searched the archives for X-ray imaging. This is useful to check for the presence of AGN
activity in an unbiased fashion, which is widely believed to be related to quenching and feedback in
galaxy formation models. All of the 8 galaxies with HST imaging plus two more have X-ray data
from the ACIS-I camera on board the {\it Chandra} X-ray telescope. We chose this as our final sample of
objects, as it spanned the widest range in luminosities and was mosu suitable for our purposes.

We then searched the archives for further pointed observations of these objects and found {\it Spitzer} imaging 
at 3.6, 4.5, 6.7 and 8.0 $\mu$m from IRAC and data at 24, 70 and 160 $\mu$m from MIPS in the Spitzer Heritage
Archive. The main properties of these 10 objects are shown in Table~\ref{properties}, showing that
they span a comparatively wide range of luminosities, stellar masses and internal velocity dispersions, 
consistent with moderately massive galaxies. Here, internal velocity dispersions are provided by the 
SDSS, while morphologies are taken from \cite{Huertas2011}: stellar masses were calculated in \cite{Melnick2013} 
as part of our analysis.

\begin{table*}
  \caption{Fundamental information on the PSGs galaxies analysed in this paper}
  \centering
  \begin{tabular}[h]{ccccccc}
  \hline
    RA (2000) & Dec (2000) & SDSS $r$ (AB)  & Redshift & $\log (M_*/M_{\odot})$ & Morphology & $\sigma$ (km s$^{-1}$)\\
    \hline \hline
    00 44 59.23 & $-08$ 53 23.0 & 13.72 & 0.0194 & 10.1 & S0  & $103 \pm 3$ \\
    03 16 54.89 & $-00$ 02 31.2 & 14.92 & 0.0232 & 10.0 & S0 & $ 89 \pm 4$ \\
    08 27 01.39 & +21 42 24.4    & 13.89 & 0.0154 & 10.0 & Sab & $77 \pm 3$ \\
    09 03 32.77 & +01 12 36.4    & 16.23 & 0.0580 & 10.2 & Scd & $118 \pm 5$ \\
    09 44 26.95 & +04 29 56.8    & 15.35 & 0.0467 & 10.5 & E  & $136 \pm 7$ \\
    12 39 36.03 & +12 26 20.0    & 14.55 & 0.0409 & 10.3 & S0   &  $185 \pm 6$ \\
    13 05 25.82 & +53 35 30.3    & 14.40 & 0.0380 & 10.5 & E/S0  & $142 \pm 4$  \\
    16 13 30.18 & +51 03 35.6    & 15.54 & 0.0338 &  8.7  & Amorphous & ...  \\
    16 27 02.55 & +43 28 33.9    & 14.45 & 0.0462 & 10.5 & E & $171 \pm 5$ \\
    22 55 06.80 & +00 58 40.0    & 14.91 & 0.0532 & 10.5 & E &  $187 \pm 7$\\
   \hline
   \end{tabular} 
   \label{properties}
\end{table*}

We discuss the appearance and colour gradients of these objects in detail below. These are generally 
early-type galaxies (mostly E/S0 with some early-type spirals), having luminosities and stellar masses 
typical of approximately $L^*$ galaxies (\citealt{McMillan2011} estimates the Milky Way stellar
mass at $\log M_*/M_{\odot} \sim 10.8$) and colours between the blue cloud and red sequence. The 
images show that our galaxies are observed in various stages of interactions: from close pairs (SDSS
J0903+0112 and SDSS 1613+5103, where the latter has prominent tidal features), to obviously disturbed
merger remnants (SDSS J0944+0429, J1239+1226, J1627+4328, 2255+0058) and comparatively
relaxed galaxies such as SDSS J0316--0002, J0827+2142 and 1305+5335 (and possibly J0044--0853).
Many of these systems are visibly asymmetric and show prominent, high surface brightness features
such as shells and tails, visible even in the shallow SDSS images. Simulations by \cite{Conselice2006}
on the evolution of galaxy asymmetries as a function of time, indicate that these galaxies must have
interacted or merged within the last $\sim 200$ Myr. 

\section{Morphologies and Colours}

For all galaxies with HST data, we examine the images in detail, derive the $r$-band surface brightness
profile and the $B-r$ colour distribution as a function of radius (for SDSS J1613+5103 we also consider
the $U-B$ radial profile as well), and model them with a single Sersic profile using the GALFIT software
\citep{Peng2002,Peng2010}. We also examine the residuals after the smooth model is subtracted. All
derived parameters are tabulated in Table~\ref{galfit}. Given the complexity of most of these objects, not
all residual maps are fully satisfactory, but it is difficult to model such asymmetric and amorphous
galaxies and achieve perfect subtraction.

\begin{figure}
\centering
\includegraphics[width=0.45\textwidth]{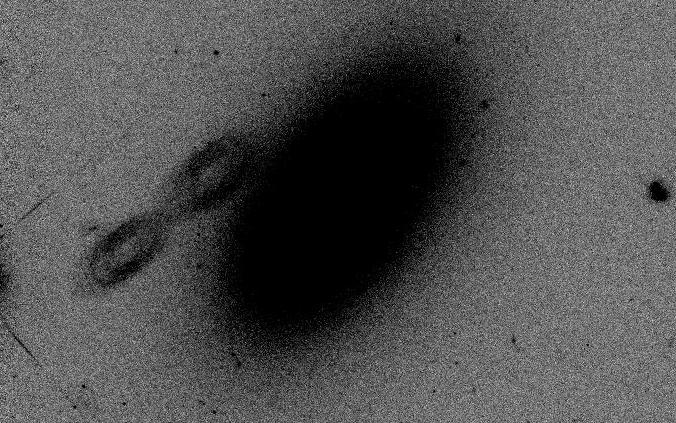}\\
\includegraphics[width=0.45\textwidth]{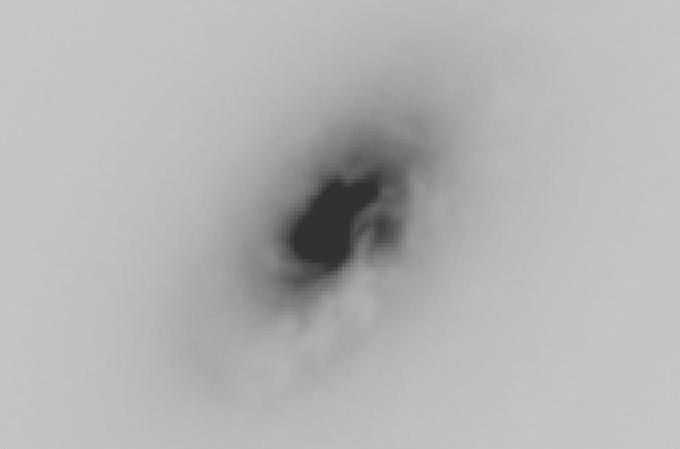}\\
\includegraphics[width=0.45\textwidth]{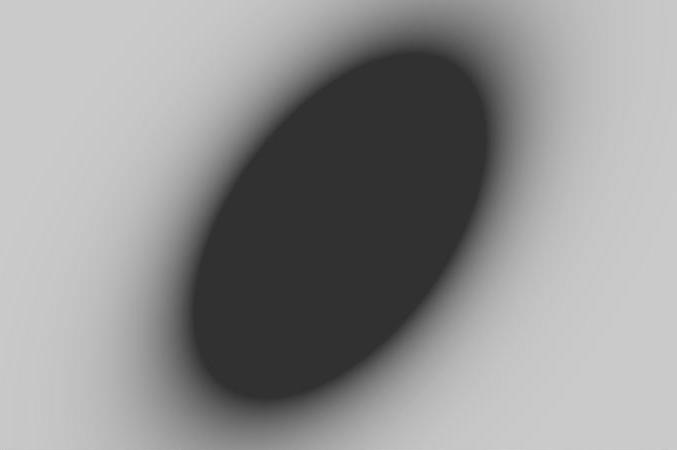} \\
\includegraphics[width=0.45\textwidth]{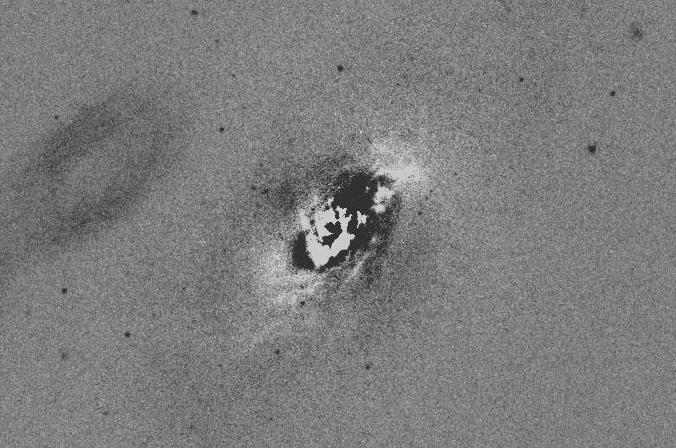}\\
\caption{HST images in the $r_{F625W}$ band for SDSS J031654.00$-$000231.4. On top, 
the original data from the WFC3 camera. Second from top, a zoomed image to the galaxy centre, 
with the appropriate stretch to reveal the inner features, especially the central dusty spiral in this 
case. Second from bottom: the best GALFIT model. Bottom: the residuals after the model is 
subtracted from the galaxy image. Similar data for all other galaxies with HST imaging are 
presented in Appendix A.}
\label{j0316img}
\end{figure}

{\it SDSS J0044--0853} We do not have HST imaging for this object, but the available SDSS data
shows a comparatively normal early-type galaxy, with little evidence of morphological disturbance.

{\it SDSS  J0316--0002} In the SDSS and HST imaging this object (Markarian 1074) appears as a
normal early-type spiral. However, in the core region we observe a bright central nucleus and an
internal dusty spiral (Fig.~\ref{j0316img}). After removal of the GALFIT model, the residual image
(also shown in Fig.~\ref{j0316img}) shows weak spiral structures and the complex central features
we have remarked upon. The core is actually red, but it becomes rapidly bluer out to $\sim 1''$
(Fig.~\ref{j0316prof} and afterwards the $B-r$ colour reddens out to $\sim 10''$. Positive
colour gradients are typical of PSGs \citep{Yang2008}. The rapid blueing in the inner $1''$ is
probably due to young stellar populations: keeping in mind the dustiness of the core, this may
imply that star formation was quenched outward. We plot similar figures for all other galaxies in
the study in the on-line Appendix.

\begin{figure}
\centering
\includegraphics[width=0.45\textwidth]{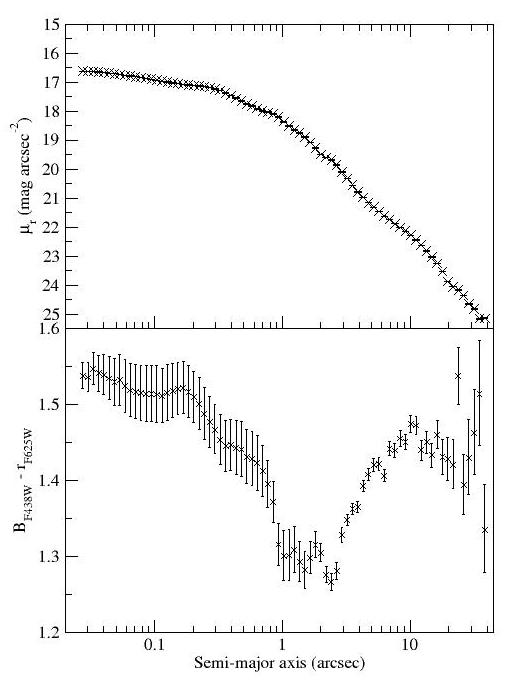}
\caption{Top: The $r_{F625W}$ band surface brightness profile (Vega magnitudes) for J0316--0002. Bottom:
The $B_{F438W}-r_{F625W}$ radial colour distribution. The equivalent data for all other systems are in Appendix
A.}
\end{figure}
\label{j0316prof}

{\it SDSS J0827+2142} This galaxy appears to be an edge-on S0/a. The inner region contains a bright
nucleus and evidence of a dust lane. This is also seen in the residuals of the GALFIT model (Fig.~\ref{j0827img}). 
As with J0316--0002, the core of the galaxy is red but it becomes rapidly blue 
out to $1"$ and is then positive (reddening outwards) beyond (Fig.~\ref{j0827prof}).

{\it SDSS J0903+0112} We also do not have HST images for this object. However, the SDSS data
show a very close pair of galaxies, with what may be a tidal envelope or large tail. This appears to
be an imminent merger, with luminosity ratio of $\sim 1:4$. The brighter galaxy is the PSG, while the
fainter object is quiescent. This suggests that an increase in star formation has been triggered by the 
interaction from pre-existing gas.

{\it SDSS J0944+0429} An obvious merger remnant, with a large shell, a plume and numerous
dust lanes, resembling Cen A in structure. The GALFIT model removes the smooth structure and
allows us to better see the high surface brightness residuals due to the merger (Fig.~\ref{j0944img}),
including a bright central cusp. This galaxy has a very blue core, reddening outwards in the inner $1''$ 
and then resuming a more normal blueing (negative gradient) trend out to $10''$, although the inner $0.2"$
may be slightly redder (Fig.~\ref{j0944prof}); interpretation of these trends may be complicated by the 
dust lanes, but they are consistent with recent star formation in the centre and outward quenching. The 
outer regions may be dominated by the shell structure or the previous (negative) gradients of the host 
galaxy.

{\it SDSS J1239+1226} An obvious merger remnant with prominent shells and plumes. The main
body of the galaxy has somewhat boxy isophotes. We also see a bright central nucleus. GALFIT
subtracts the smooth component well, revealing more detail in the tidal features (Fig.~\ref{j1239img}. 
There are several point sources which are likely super star clusters \citep{Yang2008}. The inner
$0.3"$ of this galaxy are red, with a rapid blueing out to $1"$, similar to J0316--0002; the colour
gradient out to $30"$ is positive, becoming rapidly red outwards (Fig.~\ref{j1239prof}).

{\it SDSS J1305+5330} A more normal object (Markarian 239), it does not seem to have a
compact nucleus. Models and residuals show evidence of weak spiral arms and possibly 
a small central bar (Fig.~\ref{j1305img}). As with J1239+1226, this galaxy has an inner red
core, blueing out to $0.3"$ and then a positive gradient outwards, with several features probably
due to the spiral arms (Fig.~\ref{j1305prof}).

{\it SDSS J1613+5103} A prominent interacting pair, already studied in detail by \cite{Yagi2006}
and \cite{Goto2008b}. The PSG has an amorphous morphology, with a large tidal plume and a
faint bridge connecting it to the brighter (and quiescent) galaxy, which is also tidally disturbed. 
Fig.~\ref{j1613img} shows that the PSG hosts numerous super star-clusters, many of which 
appear to be along 'jets' emerging from the PSG core. We discuss these below. GALFIT does
not return a useable fit to the galaxy light, although it seems to indicate a very high Sersic index
consistent with a bulge-dominated object or peaked light distribution. The core of this object is
very blue (Fig.~\ref{j1613prof}), although the inner $0.2"$ are somewhat redder in $U-B$. There
is a flat gradient indicating a homogeneously young population in the inner $1"$ followed by a
postive gradient (red outwards) to $\sim 10''$. This is consistent with a central starburst. \cite{
Goto2008a} argues that star formation in this galaxy must have proceeded somewhat longer in
its outskirts and be quenched from the core out, which is broadly consistent with the colour
gradients derived from our higher resolution data.

{\it SDSS J1627+4328} Fig.~\ref{j1627img} shows this galaxy to be a highly disturbed galaxy, with
numerous tidal arcs, shells, plumes and fans. There is a bright central nucleus surrounded
by shell and dust lanes. The GALFIT profile models this galaxy well and reveals numerous
more low surface brightness features. The galaxy has a red core in Fig.~\ref{j1627prof} that
becomes rapidly blue to $2''$ (as in J0316--0002 and other galaxies). The gradient to $\sim 
10''$ is positive as in many other PSGs in this and other samples.

{\it SDSS J2255+0058} There is an obvious large shell surrounding this galaxy in Fig.~\ref{j2255img}
as well as a tidal cloud and two 'jets' emerging from the core and containing super star clusters. The
GALFIT residual images reveal a rich ensemble of tidal streams and arcs. This is the only one of our
objects with a $\sim 3.9 \sigma$ detection in the Chandra Source Catalog \citep{Evans2010}: see
below for a study of the X-ray properties of our galaxies. There is a blue core in Fig.~\ref{j2255prof},
although it may become somewhat redder in the inner $2''$. At large radii the gradient is again positive
which is consistent with previous work and other galaxies in this sample.

\begin{table}
  \caption{GALFIT parameters for Single Sersic profiles}
  \centering
  \begin{tabular}[h]{cccc}
  \hline
    Object & Model $r$ (AB)  & $R_e$ (arcsec) & $n$ \\
    \hline \hline
    J0044$-08$53 & ...                              & ... & ... \\
    J0316$-00$02 & $14.89 \pm 0.01$ & $ 6.93 \pm 0.01 $ & $ 3.96 \pm 0.01 $ \\
    J0827+2142   & $14.17 \pm 0.01$ & $ 5.00 \pm 0.01 $ & $ 1.26 \pm 0.01 $ \\
    J0903+0112  & ...                              & ...                             & ...      \\
    J0944+0429 & $15.57 \pm 0.01$  & $4.92 \pm 0.01$ & $2.93 \pm 0.01 $ \\
    J1239+1226 & $14.00 \pm 0.01$ & $22.67 \pm 0.09$ & $8.46 \pm 0.01$ \\
    J1305+5335 & $14.33 \pm 0.01$ & $ 3.16 \pm 0.03$ & $5.52 \pm 0.01$ \\
    J1613+5103  &  ... & ... & ... \\
    J1627+4328  & $14.49 \pm 0.01$  & $ 6.41 \pm 0.01$ & $3.19 \pm 0.01$ \\
    J2255+0058 & $14.92 \pm 0.01$ & $ 9.72 \pm 0.03$ & $3.37 \pm 0.01$ \\
   \hline
   \end{tabular} 
   \label{galfit}
\end{table}

\section{Stellar Populations}

\begin{figure}
\centering
\includegraphics[width=0.5\textwidth]{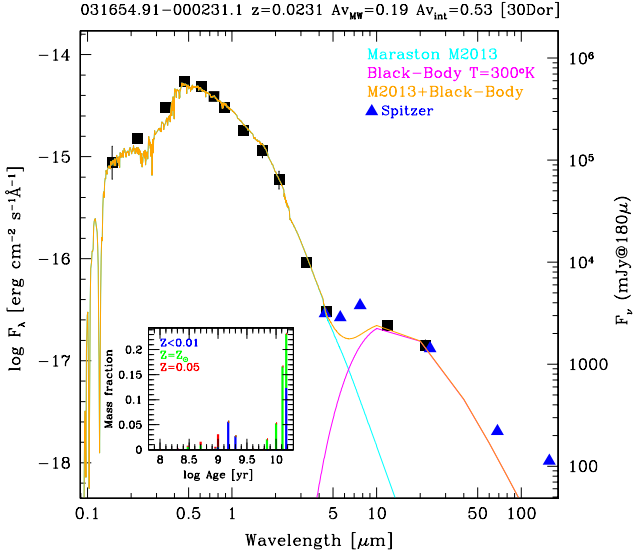}\\
\caption{Spectral energy distribution, SDSS spectrum and  Maraston M2013 model
fit for J0316--0002 shown by the cyan line. The magenta line corresponds to an ad-hoc $T_{dust}=300^o$K black-body 
component added to the model SEDs to fit the MIR data. The inset shows the distribution of stellar populations in ages and metallicity given 
by the {\sc Starlight} code and used to compute the synthetic SEDs. The {\sc Spitzer} (blue triangles) data illustrate the presence of a cold-dust 
component in addition to the 300K component used to fit the SED. }
\label{sed}
\end{figure}

We have modelled the SDSS spectra with {\sc Starlight} \citep{Cid2005} to reconstruct the
star formation history of these objects. This uses the latest version of the \cite{BC03}
spectral library with 24 age and 5 metal abundance bins. The star-formation histories resulting from the 
{\sc Starlight} models were then used to construct synthetic Spectral Energy Distributions (SED) using 
the latest version of the \cite{Maraston2005} models -- M2013 -- as described in \cite{Noel2012} and \cite{Melnick2014}.
We compare the synthetic SEDs, for which the only free parameter is the photometric zero-point, 
anchored to the SDSS $i$-band, with the available photometry from 0.2 to 160 $\mu$m. Fig.~\ref{sed} shows
 the results of this process for J0316--0002; similar figures for all other galaxies can be found in the Appendix. 
In each of these figures an inset shows the star-formation histories of each galaxy from {\sc Starlight} for three 
broad metallicity bins (centred on 1/2 solar, solar and twice solar abundance)

Figure~\ref{sed} illustrates that the updated \cite{Maraston2005} models yield a very good fit to the spectral energy distributions,
from the $u$ to the $4.5$ $\mu$m bands probing stellar emission, but not always in the FUV and NUV Galex bands. At $\lambda>5\mu$m 
we observe an excess above the stellar blackbody tail, consistent with the presence of two dust components: the warm component at 
$\sim 300^{\circ}$K discussed above and a much colder component responsible for the flux at $\lambda>20\mu$m detected by {\sc Spitzer}. 
The warm component is likely heated by the young stellar populations and Asymptotic Giant Branch stars \citep{Chisari2012} which also
power LINER emission as well. \cite{Maraston2005} does not model explicitly dust emission from AGB stars so the discrepancy is not
surprising. Despite the poor resolution of mid-infrared data, most objects are reasonably well resolved in WISE or Spitzer images, so that
no central point source appears to be responsible for the $\lambda > 5 \mu$m emission. Our mid-infrared fluxes are generally lower than
those of weak AGNs or starburst galaxies in the sample by \cite{Lutz2004}.

\begin{figure}
\centering
\includegraphics[width=0.5\textwidth]{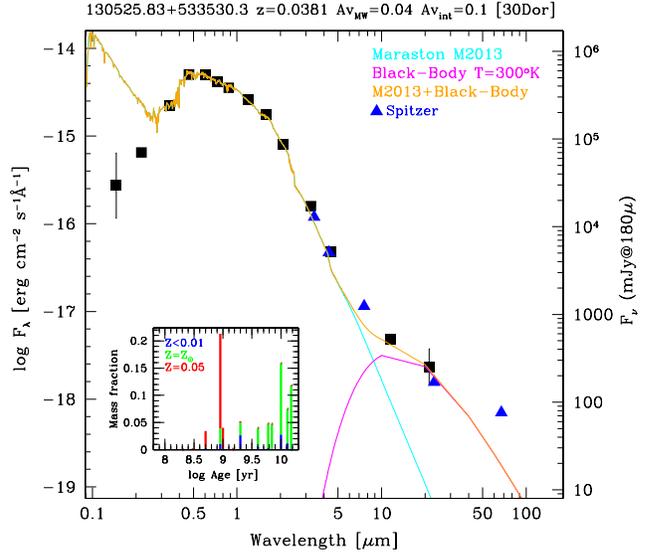}
\caption{Same as Fig.~\ref{sed} but for J1305+5335, where we observe a strong deficit
of NUV and FUV light compared to the models fitted to the optical and infrared light.}
\label{sed2}
\end{figure}

However, we also see that, on the basis of the stellar populations derived from the spectra -- within the Sloan $3"$ fiber aperture -- 
the models conspicuously over-predict the spectral energy distribution in the NUV and FUV, often by large factors.
As an example we show below the SED for J1305+5335 in Fig.~\ref{sed2}, which shows a dramatic deficit in UV
flux compared to the models fitted to the optical and the spectra.  This 
is difficult to explain even for dramatically different extinction curves in the far ultraviolet. We generally do
not see this excess for our large sample of PSGs (at a higher mean redshift) discussed in \cite{Melnick2013}. 
One possibility is that the $6"$ apertures used by GALEX contain much less flux than extrapolated from the
stellar populations in the $3"$ fiber aperture of SDSS spectra and this is an aperture effect. This may
be consistent with the UV-bright (young) populations being restricted largely to the galaxy cores, as 
argued by the observations that the inner $\sim 1"-2"$ in these galaxies tend to be bluer and 
probably contain the majority of the younger stellar populations.

The stellar populations are consistent with what we would naively expect from the morphologies of these
objects. In SDSS J0044--0853 the dominant population is old (10 Gyr) and approximately solar; we also
observe a $\sim 10^9$ year old super-solar component contributing about 20\% of the present-day stellar 
mass (Fig.~\ref{j0044sed}). The latter was likely formed during the most recent star formation episode, likely
from pre-existing gas which had been processed by the main galaxy. SDSS J0316--0002 (Fig.~\ref{sed})
has an old solar and sub-solar population and a younger ($\sim 1$ Gyr) component, increasing in metal abundance with 
decreasing age, arguing for a prolonged star formation episode with self-enrichment. In
J0827+2142 (Fig.~\ref{j0827sed}) we see an older solar-metallicity component ($\sim 10$ Gyr), followed 
by a more prolonged star formation episode lasting a few Gyrs and a more recent starburst that produced
super-solar stars, consistent with an enhanced star formation rate fuelled by pre-existing gas in the galaxy
as in J0044--0853. In J1305+5330, Fig.~\ref{sed2} shows an old ($\sim 10$ Gyr) sub-solar and solar
component. The galaxy seems to have continued to form stars at moderate rates until $\sim 1$ Gyr ago, 
where we see the onset of a metal-rich stellar population formed in a starburst and the termination of all
star formation a few $10^8$ years ago. All of these galaxies have comparatively relaxed appearances in 
Figures~\ref{j0316img},~\ref{j0827img} and ~\ref{j1305img}, as well as in the available SDSS image for
J0044--0853.

The disturbed remnant J0944+0429 appears to be a largely old galaxy which formed the majority of its
stars about 10 Gyrs ago, but has now undergone a large starburst, a few hundreds Myr ago, comprising
$\sim 10\%$ of its present mass. The new component is metal-poorer and this suggests that this object
may have undergone a merger with a low mass gas-rich galaxy, similar to other dusty ellipticals (e.g.,
\citealt{Kaviraj2013}). J1239+1226 has a dominant metal-rich population (Fig.~\ref{j1239sed}) with an
age of 10 Gyr, but also contains a few 10$^8$ year old metal-rich component, which was likely formed
during the merger from pre-enriched gas. In SDSS J1613+5103 we observe a young ($\sim 10$ Myr) and 
metal-poor population, which was likely produced during the current interaction. There are also older 
metal-poor stars (a few Gyrs old) and intermediate-age solar and super-solar components that may 
have been formed during a previous close passage. The low metallicity of the younger population 
suggests that metals formed in the previous star formation episode were not efficiently retained or t
that the galaxy has accreted more pristine gas, triggering star formation at a later time (Fig.~\ref{j1613sed}). 
For J1627+4328, the main population is old and metal-rich, with evidence for self-enrichment, but we also 
see a small young and metal poor component $\sim 1$ Gyr old (Fig.~\ref{j1627sed}). This is likely to have
originated in a merger with a gas-rich low-mass  galaxy as well. Finally, Fig.~\ref{j2255sed} shows the 
spectral energy distribution for J2255+0058. This shows a similar pattern with old stars and younger
populations a few 10$^8$ years old; however, these show a mixture of abundances and one possibility 
is that fuel was contributed by the massive galaxies as well as a more metal-poor companion.

\subsection{The super star clusters of J1613+5103}

In SDSS J1613+5103 we also observe numerous super star clusters spread across the face of the galaxy,
with several bright objects apparently lying among jets emerging from the main body of the PSG. This
`cluster stream' is clearly visible in the image of the central part of the galaxy shown in Figure~\ref{j1613img}. 
Since we have $U$, $B$ and $r$-band data we can attempt to model the star formation histories of these 
clusters. These provide a secondary check on the ages derived from fitting stellar population models to the
integrated spectra, at least for the younger stellar populations. They are also some of the most distant such
clusters yet observed.

\begin{figure}
\centering
\includegraphics[width=0.45\textwidth]{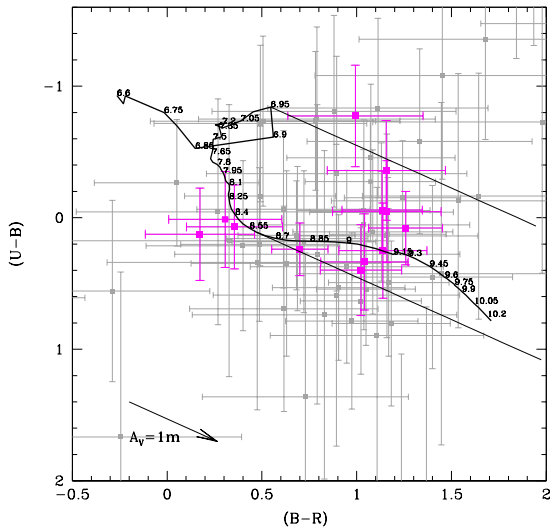}
\caption{Colour-colour diagram for super star clusters in the field of SDSS J1613+5103. The
thick line represents the age sequence of single-burst solar metallicity cluster from Girardi et
al. (2000) and Marigo et al. (2008), with ages labelled in the plot logarithmically in Myrs. The grey 
points represent all the clusters for which we have photometry, while the magenta points show clusters 
with photometric errors of less than 0.5 mag in each color.  The reddening vectors are shown by the solid 
lines and by the arrow in the bottom left corner of the plot. }
\label{ccd}
\end{figure}

We have selected all point sources within $30''$ of the PSGs as candidate clusters and have then
measured aperture magnitudes for these objects using standard routines. The resulting color-color diagram 
for these clusters (for $r < 28$) is presented Fig.~\ref{ccd} together with the age sequence sequence for 
single-burst clusters of solar metallicity from the latest Padova isochrones \cite{Girardi2000,Marigo2008}. 

As mentioned above this galaxy is very dusty so the dust extinction varies significantly from cluster to 
cluster in this diagram. Therefore we are not able to assign reliable ages to each cluster. Clearly, 
however, if we restrict our attention to the best observed objects (shown in colour), we see that all 
have ages between a few million years and about one Gyr, with a peak about 10$^8$ years ago. 
This is largely consistent with the star formation history inferred from analysis of the spectrum, suggesting
that our analysis is appropriate. Super star clusters are known to exist in many PSGs, including those studied
by \cite{Yang2008} and in J1239+1226 and J2255+0058 in our sample: however, in these cases we have no 
$U$ band data and we cannot carry out a similar analysis, as we are not able to estimate a reddening vector. 

\section{Active Nuclei in PSGs}

Feedback from an AGN could provide the means to truncate star formation in a massive galaxy.
As many of our galaxies are obvious merger remnants (whose high visibility phase is expected
to last for $\sim 200$ Myr -- Conselice 2006), star formation appears to have been centrally 
concentrated (as observed in other PSGs) and to have been suppressed outwards (e.g.,
\citealt{Goto2008b}) over relatively short periods (a few 10$^8$ years from analysis of the
stellar populations and as also found by \citealt{Brown2009}), we might suppose that AGNs
have played some part in the origin of the PSG signature. If the AGN was triggered during the
final coalescence and shortly after the major star formation episodes (as in simulations by
\citealt{Springel2005} and \citealt{Hopkins2008a} -- note that \citealt{Bessiere2014} finds that
in the Type II QSO J002531--104022 star formation activity must be nearly simultaneous with
the AGN), we might be able to detect even obscured young AGNs in their cores down to a small 
fraction of the Eddington luminosity \citep{Hopkins2009b}. In these models, AGNs are initially
deeply buried by gas and dust and only become optically visible a few hundred Myr after the
peak of star formation activity, approximately the age of the young stellar populations in our
objects. However, they should be visible in the X-rays or radio even if they are highly obscured
in the optical.

\begin{figure}
\centering
\includegraphics[width=0.45\textwidth]{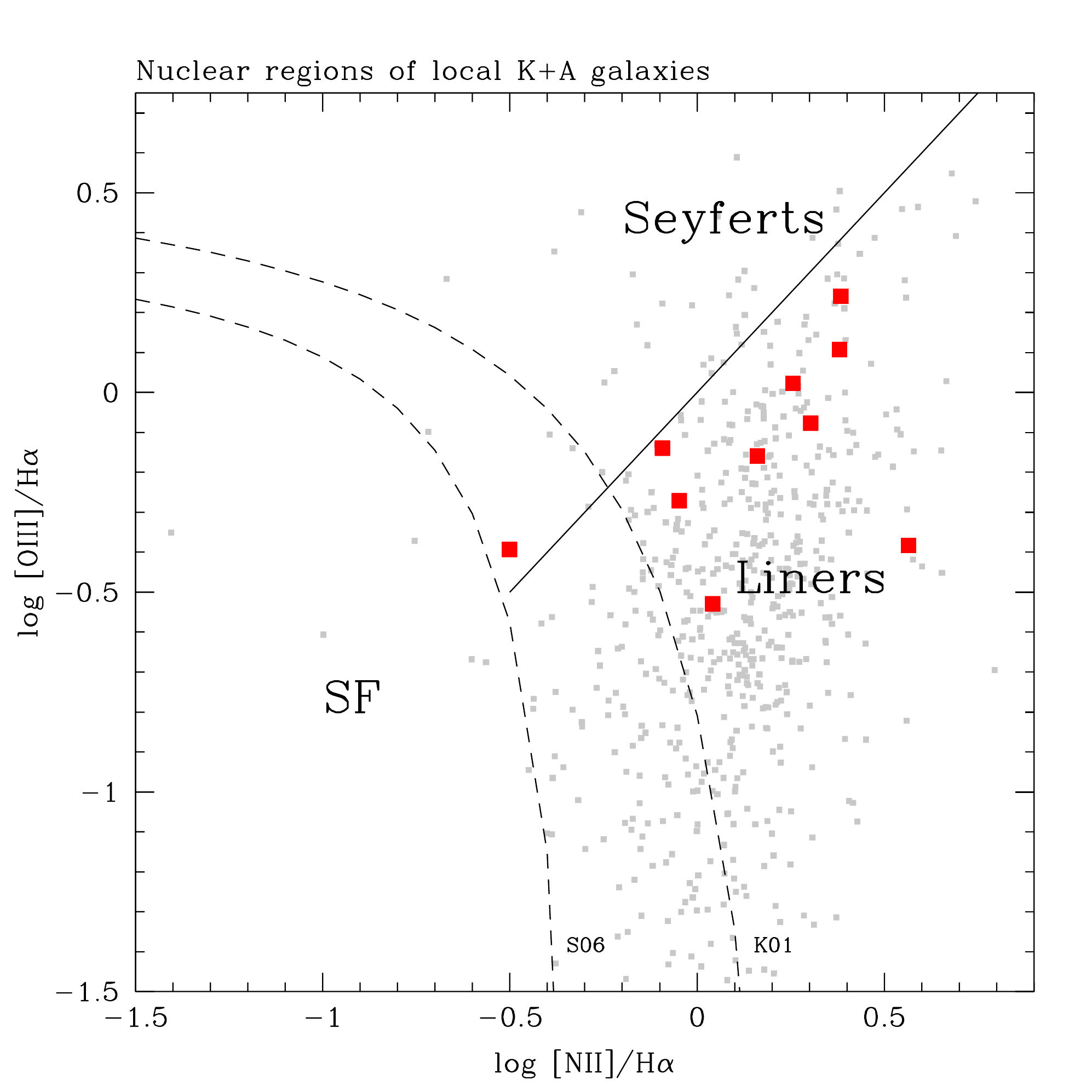}
\caption{Line ratio diagram for all 808 K+A galaxies (grey points) in Melnick \& De Propris (2013)  and 
the 10 galaxies studied in this paper (red squares). We show the regions inhabited by star-forming galaxies, 
LINERS and AGN hosts (Kewley et al. 2001, Stasinska et al. 2006). None of our objects falls among the 
Seyferts.}
\label{bpt}
\end{figure}

We first plot the conventional line-ratio diagram, which separates AGNs from LINERS and
star-forming galaxies \citep{Kewley2001,Stasinska2006} in Fig.~\ref{bpt}. None of our galaxies
falls among the Seyferts or Star-formers. They are all among the LINERS, as do most of the
PSGs in \cite{Melnick2013}. However, the majority of LINERS are not weak AGNs, but are
powered by young stellar populations and/or ultraviolet-bright AGB and post-AGB stars
\citep{Stasinska2008}: this is indeed the case for PSGs, very few of which fall in the
weak AGN section in \cite{Melnick2013}. 

\begin{figure}
\centering
\hspace{-0.5cm} \includegraphics[width=0.45\textwidth]{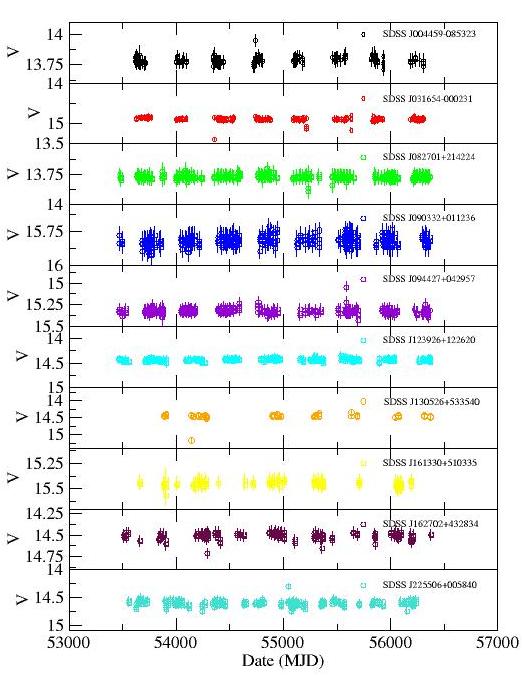}
\caption{Variability of our objects (indicated in the figure legend) in the $V$ band from the Catalina Rapid Transient
Survey photometry}
\label{var}
\end{figure}

AGNs often show optical variability \citep{Hawkins2002}. Our galaxies lie within the footprint of
the Catalina Rapid Transient Survey \citep{Drake2009} and we plot their Catalina $V$ magnitude
vs. modified Julian Date in Fig.~\ref{var}. None of the galaxies studied in this paper shows signs
of optical variability.

\begin{figure}
\centering
 \includegraphics[width=0.4\textwidth]{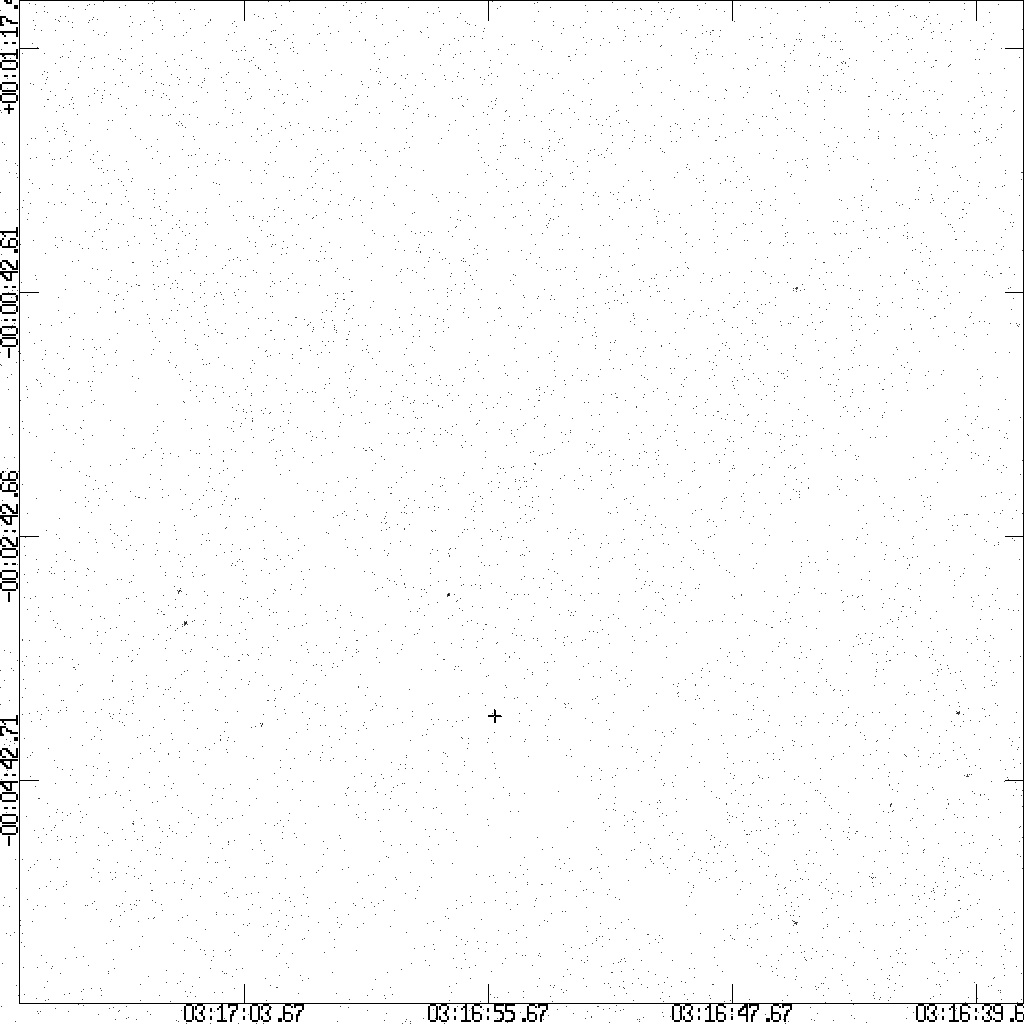}
\caption{{\it Chandra} ACIS-I image of J0316--0002 (central chip). The cross shows the position of the
galaxy. No clear detection can be found.}
\label{X-rays}
\end{figure}

X-ray emission unambiguously identifies the presence of AGNs, especially above 2 KeV \citep{Brandt2005}.
We have used archival {\it Chandra} images, taken with the ACIS-I camera with exposure times of 10 to 20 ks,
to search for X-ray emission in the 0.5--7.0 KeV band. In Fig.~\ref{X-rays} we show the central ACIS-I chip for
J0316--0002, with the galaxy position marked with a cross. All X-ray images are dominated by noise and show
no point source corresponding to any of our galaxies: for economy of space we do not include the other images,
but these can be easily retrieved from {\it Chandra} archive server with the dataset identifiers we provide below in
Table~\ref{xtable}. We have calculated the fluxes and upper limits using CIAO and show these in Table~\ref{xtable}.
These are consistent with absolute values of $< 10^{40}$ ergs s$^{-1}$, well below typical X-ray fluxes from 
powerful AGNs.

\begin{table}
\caption{X-ray fluxes and upper limits}
\begin{tabular}{cccc}
\hline
Object & Dataset & Flux & upper limit \\
           &   {\small ADS/Sa.CXO\#obs/ }           &        [ergs/s]            &  1$\sigma$ [ergs/s] \\
\hline
\hline
J0044--0853 & 10266  & $2.1 \times 10^{38}$ & $2.1 \times 10^{39}$ \\
J0316--0002 & 10267  & ...                                   & $7.9 \times 10^{39}$ \\
J0827+2142 & 10268  & ...                                   & $8.8 \times 10^{38}$ \\
J0903+0112 &   8174 & $6.9 \times 10^{39}$ & $2.7 \times 10^{40}$ \\
J0944+0429 & 10269  & ...                                   & $ 5.1 \times 10^{39}$ \\
J1239+1226 & 10270 & $7.0 \times 10^{39}$ & $1.6 \times 10^{40}$ \\
J1305+5335 & 10271  & $6.2 \times 10^{39}$ & $2.4 \times 10^{40}$ \\
J1613+5103 & 10272  & $5.1 \times 10^{40}$ & $7.8 \times 10^{40}$ \\
J1627+4328 & 10273  & $1.8 \times 10^{40}$ & $4.6 \times 10^{40}$ \\
J2255+0058 & 9594  & $2.1 \times 10^{40}$ & $4.9 \times 10^{40}$ \\
\hline
\end{tabular}
\label{xtable}
\end{table}

If we then take the central velocity dispersions in Table~\ref{properties} and calculate the expected 
black hole masses following the relation shown by \cite{Graham2013}, we can use the X-ray fluxes 
above to estimate that no black hole can be present in these galaxies with an accretion rate greater
than about 0.1\% of the Eddington luminosity. The X-ray fluxes and upper limits are more consistent 
with the integrated flux from X-ray binaries in these comparatively young populations
\citep{Gilfanov2004}. Supermassive black holes are expected to be detectable at these luminosities 
for a period of $\sim 1$ Gyr after they are activated \citep{Hopkins2009b}, even if they have already
quenched star formation and stopped accreting matter efficiently.

Furthermore, we also use the BAT/Swift 70 month map \citep{Baumgartner2013} to estimate the
likelihood that a hard X-ray source is present within the $2.8'$ pixel containing these galaxies. Only
in one case we have signal to noise greater than 3 (this is next to an unrelated AGN, whose PSF may
spread into the pixel of interest) and most have signal to noise well below 1. 

\begin{figure}
\centering
\includegraphics[width=0.45\textwidth]{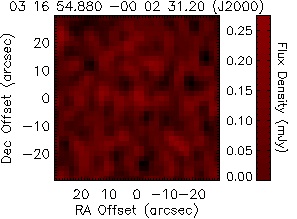} 
\caption{Cutout of data from the FIRST survey centred on J0316--0002. The typical rms noise per beam 
is 0.15 mJy, for a $5\sigma$ detection limit of 0.75 mJy per beam.  The image is $1'$ on the side. No source 
is found.}
\label{radio}
\end{figure}

We also use the FIRST survey to estimate the 20cm radio flux at these positions. A cutout of the
FIRST image for J0316--0002 is shown in Fig.~\ref{radio}. All other galaxies are similar: no point
source is present at any of the stated positions (the images can be easily retrieved from the FIRST
survey and are not presented here). Since the $5\sigma$  sensitivity of the FIRST survey is 0.75 mJy per beam,
this corresponds to absolute radio powers $< 10^{23}$ W Hz$^{-1}$ for all our systems. Following \cite{Sadler2002} 
none of these galaxies is likely to host a radio source powered by an AGN.

We conclude that there is no strong evidence for the presence of AGNs in these galaxies. In agreement 
with \cite{Shin2011}, current AGN activity seems to be unrelated to the PSG phenomenon.

\section{Discussion}

We have studied 10 PSGs with a wealth of imaging in multiple bandpasses and archival data in the
X-rays and radio, as well as spectroscopy. Two of these galaxies are interacting pairs, at least one
showing large tidal disturbances (J1613+5103). At least four more are obvious merger remnants,
with numerous high surface brightness tidal features (shells, fans, plumes, etc.) sometimes visible
even in the shallower SDSS images. Four more of these galaxies are comparatively normal systems,
but even in these cases there is some evidence of an old interaction (e.g., the dusty spiral in J0316--0002).

All of these galaxies have old stellar populations but they also contain significant young components,
with ages ranging between a few 10$^7$ and 10$^9$ years. There is of course considerable variation
from galaxy to galaxy. All show evidence of an old ($\sim 10^{10}$) initial star formation episode, 
consistent with all galaxies of similar masses. In many cases we then see long periods of quiescence,
although some galaxies show evidence of prolonged star formation in the past (we do not address
J0903+0112, where the stellar populations seem to have been formed simultaneously $\sim 1$ Gyr
ago, that seems little consistent with the apparently normal morphologies and colours -- we attribute
this to contamination in the fiber that may have been placed to the geometric center of these two
close systems). The recent starburst shows a variety of ages and metallicities, and this implies that
the nature of the available fuel and of the interaction have been different from case to case, as one
expects from 'stochastic' mergers. In some cases the young stars are metal rich and may have been
produced by a dramatic increase in star formation that used up the available fuel in an old galaxy. In
others, the young stars are more metal poor and this may imply that additional gas has been accreted
from a lower-mass companion or that the original gas has been diluted by infall of more
pristine material. Several PSGs (albeit, none of the ones we study) have been found to contain 
significant amounts of HI gas, comparable to those found in spiral galaxies of similar luminosities
\citep{Chang2001,Buyle2006,Zwaan2013}. \cite{Nielsen2012} also argued that we may be witnessing
galaxies in different portions of an activity cycle, with several star formation episodes separated by a
few Gyrs. This is supported by the star formation history of SDSS J1613+5103, for example.

Although it is generally difficult to date a merger event based on morphological indicators, as this
depends on the nature of the merger, the gaseous content and the detailed mechanics of the 
encounter, more asymmetric galaxies are expected to be observed closer to the epoch of final
coalescence \citep{Conselice2006}. In our case, there is a broad relation between the age of the
stellar populations and their contribution to the galaxy mass, and the degree of morphological
disturbance. This is similar to earlier observations by \cite{Schweizer1992} and more recently
by \cite{Gyory2010} and suggests that the mergers and interactions have played a role in 
triggering or increasing star formation in these galaxies. However. based purely on this sample,
we would argue that the peak of star formation precedes the epoch of closest approach and
coalescence, and may instead take place during earlier close passages. This is clearly the case
in J1613+5103 for example. 

\cite{Yang2008} observed that most PSGs have positive colour gradients and blue cores. We
find similar results but we also observe that some galaxies have red cores in the inner $0.3"$,
or cores that become rapidly bluer outward and then start a reddening trend to large radii. It
is not straightforward to interpret these colour gradients, as the galaxies are very dusty. However,
such large colour variations as observed here are more likely to be due to age rather than metal
abundance gradients (which are usually of the order of a few tenths of magnitude per decade in
radius -- e.g., \citealt{LaBarbera2010}). 

All galaxies in the sample have blue cores compared to their outskirts. Among the more
obvious merger remnants, in J1613+5103 we see a blue core in both colours ($U-B$ and $B-r$) 
with a flat colour gradient out to $\sim1''$ (the inner 2 kpc$h^{-1}$), reddening rapidly outwards.
J0944+0429 also has a blue core that reddens outwards to $1''$ and then resumes a mild negative
gradient at larger radii (as is common in galaxies). In J1239 there is a strong positive gradient and 
a blue core, but in the inner $0.3"$ the core reddens significantly. We see a very red core in 
J1627+4328 (likely due to dust) and a positive gradient throughout J2255+0058, although this 
galaxy starts to redden in the inner $0.2"$. In the more quiescent objects such as J0316--0002,
J0827+2142 and J1305+5330, we see a red inner core ($\sim1"$) which becomes rapidly blue to 
$\sim1''$  and then reddens outwards  to $10"$. From this we can make a tentative connection 
between the 'ages' derived from the merger signatures, the stellar populations, and the broad 
shape of the colour gradients. 

Star formation appears to have been concentrated in the centre, within the inner 1--2 kpc. There is 
some evidence that quenching has proceeded outwards, so that the outer kpc in the core is younger 
(e.g., in  J1603+5103 which has just stopped forming stars we see some tentative evidence of a 
slightly redder $U-B$ colour in the inner $0.2"$). The positive colour gradients observed throughout most
of the galaxies, however, imply that star formation in these objects was also a global phenomenon,
extending across the whole galaxy. This may be compared with the results of integral field units.
\cite{Pracy2005} (albeit in a cluster sample, which may be different) and \cite{Goto2008a} reported
the existence of H$_{\delta}$ gradients in several PSGs, while \cite{Goto2008b} claimed that
there was an age gradient (somewhat older inwards) in J1603+5103. \cite{Pracy2012,Pracy2013}
argued for Balmer-line gradients in their sample and therefore for centrally concentrated star
formation, while \cite{Swinbank2012} show that the youngest population covers a significant
fraction of the galaxy surface and the starburst did not take place in the nucleus only. 

Analysis of the optical images, colour gradients and stellar populations derived from optical spectra
and multi-wavelength photometry therefore supports the view that PSGs have undergone an episode
of highly enhanced star formation following an interaction and/or merger. While this appears to have
interested a large fraction of the galaxy, it also appears that star formation was centrally concentrated.
Quenching of star formation has proceeded from the inside outwards, which is indicative of a feedback
episode (e.g., outflows as in \citealt{Tremonti2007}): the presence of bright central cusps
\citep{Hopkins2009} is also expected in this scenario.

Together with the presence of significant bulges (high Sersic indices, early-type morphological classification)
this raises the possibility that AGNs may have been important in PSGs. In simulations such as \cite{Springel2005}
and \cite{Hopkins2008b}, the merger coincides with the epoch of highest star formation and AGN activity peaks
shortly afterwards, although it is obscured for periods of a few 10$^8$ years before an optically luminous
QSO can be observed as the gas and dust are cleared. \cite{Hopkins2012} suggests looking at PSGs to observe
examples of AGNsin galaxies following merger-induced starbursts. 

However, we find little evidence that our PSGs contain any significant AGN activity, including obscured QSOs that
would be visible in the X-rays. The central black holes in these galaxies are expected to have masses of about $10^7$
$M_{\odot}$ and the X-ray luminosities (or upper limits) we measure in Table~\ref{xtable} indicate that these objects
would be accreting at well below 1\% of the Eddington luminosity. Our PSGs are observed in various stages of an
interaction, from close pairs to comparatively relaxed systems, and show a range of ages for their stellar populations.
These indicate that the peak of star formation may actually occur before final coalescence. \cite{Bessiere2012} also
finds that QSOs may occur at a variety of times, contrary to simpler modelling: \cite{Bessiere2014} actually finds 
an optically bright QSO coexisting with vigorous star formation. However, our data suggest that QSOs are not observed
(even if highly obscured) for periods of several hundred Myr after a major merger and after the ending of the major
star-forming episodes. This is similar to what has been observed by \cite{Wong2012}, where no (optical) AGNs were
found in their sample of local PSGs and by \cite{Wild2009} where no AGNs were found in their sample of $z \sim 0.8$
PSGs. On the other hand, \cite{Canalizo2013} found that the stellar populations of optically selected QSOs had intermediate
age of 0.3 -- 2.4 Gyr (somewhat larger than our systems) while \cite{Cales2013} also find examples of intermediate age
populations in optically active galaxies (but their selection includes objects with prominent [OII] emission that would not
be considered PSGs). In our data we are sensitive not only to optically visible AGNs, but also to heavily obscured systems
via our strong limits in the X-rays and radio and we can confidently state that no bright ($L > 1\%\ L_{Edd}$) nucleus is
present.

One possibility is that peak black hole accretion has taken place before the epoch at which these galaxies are observed,
and closer to the time of peak star formation; however, in the model by \cite{Hopkins2009b} such black holes should be
visible, in the X-rays, for periods of up to 1 Gyr, to the sensitivity limits we reach. On he other hand the delays between star
formation and QSO activity may be larger than the typical ages of young stellar populations in our sample, but in this case
it is difficult to understand how a much more inactive nucleus than even Sgr A, can possibly inhibit further star formation in
these galaxies. 

We can conclude that even major mergers such as some of the objects examined here, do not necessarily initiate AGN
activity and that the timescales for the QSO to accrete at high rates may be longer than expected in at least some scenarios.
In PSGs we have objects where star formation has currently ceased: yet no QSO activity is apparent to quench any residual
star formation or prevent gas (which is known to exist in these objects -- \citealt{Chang2001,Buyle2006,Zwaan2013}) from
forming new stars. Even though the sample we analyse is small, our data do not fully support a model where AGN feedback
results in post-starburst galaxies that evolve on to the red sequence.

\section*{Acknowledgments}

We highly appreciate a very helpful report from the anonymous referee.

GALEX (Galaxy Evolution Explorer) is a NASA Small Explorer, launched in April 2003. We gratefully acknowledge 
NASA's support for construction, operation, and science analysis for the GALEX mission. 

Funding for the SDSS and SDSS-II has been provided by the Alfred P. Sloan Foundation, the Participating Institutions, t
he National Science Foundation, the U.S. Department of Energy, the National Aeronautics and Space Administration, t
he Japanese Monbukagakusho, the Max Planck Society, and the Higher Education Funding Council for England. The 
SDSS Web Site is http://www.sdss.org/.

The SDSS is managed by the Astrophysical Research Consortium for the Participating Institutions. The Participating Institutions 
are the American Museum of Natural History, Astrophysical Institute Potsdam, University of Basel, University of Cambridge, 
Case Western Reserve University, University of Chicago, Drexel University, Fermilab, the Institute for Advanced Study, the 
Japan Participation Group, Johns Hopkins University, the Joint Institute for Nuclear Astrophysics, the Kavli Institute for Particle
Astrophysics and Cosmology, the Korean Scientist Group, the Chinese Academy of Sciences (LAMOST), Los Alamos National
Laboratory, the Max-Planck-Institute for Astronomy (MPIA), the Max-Planck-Institute for Astrophysics (MPA), New Mexico State 
University, Ohio State University, University of Pittsburgh, University of Portsmouth, Princeton University, the United States 
Naval Observatory, and the University of Washington.

This publication makes use of data products from the Two Micron All Sky Survey, which is a joint project of the University 
of Massachusetts and the Infrared Processing and Analysis Center/California Institute of Technology, funded by the 
National Aeronautics and Space Administration and the National Science Foundation.

This work is based in part on observations made with the Spitzer Space Telescope, obtained from the 
NASA/ IPAC Infrared Science Archive, both of which are operated by the Jet Propulsion Laboratory, 
California Institute of Technology under a contract with the National Aeronautics and Space 
Administration. 

This publication makes use of data products from the Wide-field Infrared Survey Explorer, which is a joint project of the University 
of California, Los Angeles, and the Jet Propulsion Laboratory/California Institute of Technology, funded by the National 
Aeronautics and Space Administration

The scientific results reported in this article are based to a significant degree on data obtained from the Chandra 
Data Archive. This research has made use of software provided by the Chandra X-ray Center (CXC) in the application 
packages CIAO, ChIPS, and Sherpa.

We acknowledge the use of NASA's {\it SkyView} facility (http://skyview.gsfc.nasa.gov) located at NASA Goddard Space 
Flight Center.

This research has made use of the NASA/IPAC Extragalactic Database (NED) which is operated by the Jet Propulsion Laboratory, 
California Institute of Technology, under contract with the National Aeronautics and Space Administration.

This research has made use of the NASA/ IPAC Infrared Science Archive, which is operated by the Jet Propulsion Laboratory, 
California Institute of Technology, under contract with the National Aeronautics and Space Administration. IRSA and the Spitzer Heritage 
Archive utilize technology developed for the Virtual Astronomical Observatory (VAO), funded by the National Science Foundation and the 
National Aeronautics and Space Administration under Cooperative Agreement AST-0834235.

We acknowledge the usage of the HyperLeda database (http://leda.univ-lyon1.fr).

Based on observations made with the NASA/ESA Hubble Space Telescope, obtained from the data archive at the Space Telescope Science Institute. STScI is operated by the Association of Universities for Research in Astronomy, Inc. under NASA contract NAS 5-26555.

The authors acknowledge discussions with Malcolm Bremer, Jari Kotilainen, Elina Lindfors and Steve Phillipps.

\bibliographystyle{aa}

\clearpage

\bsp

\appendix
\clearpage

\section{Imaging and surface brightness profiles for all galaxies}

Here we present the remaining images, models and residuals for all galaxies, as well as their surface brightness
profiles and colour gradients.


\begin{figure}
\centering
\includegraphics[width=0.45\textwidth]{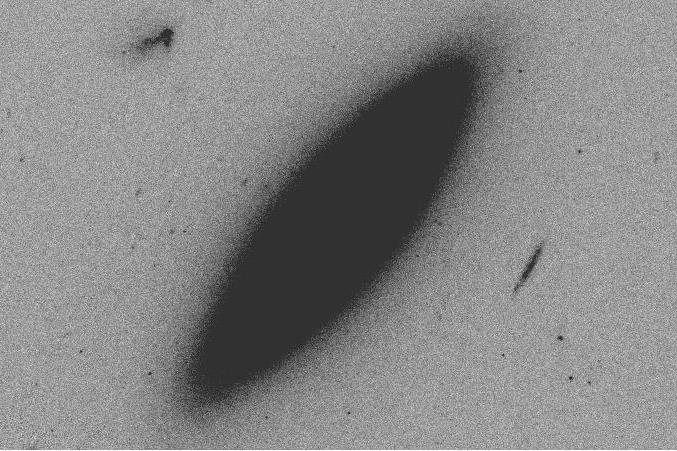}\\
\includegraphics[width=0.45\textwidth]{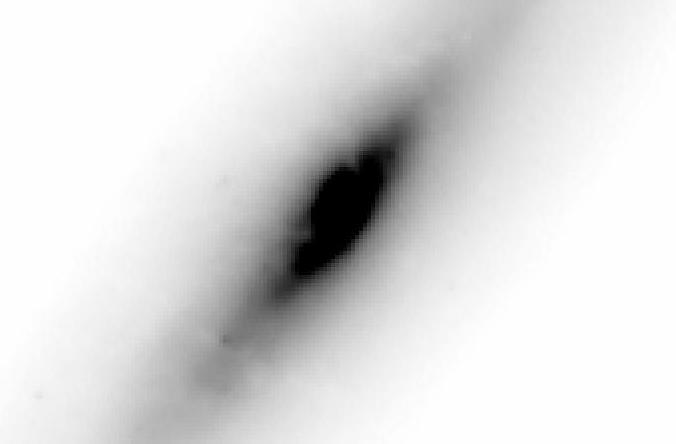}\\
\includegraphics[width=0.45\textwidth]{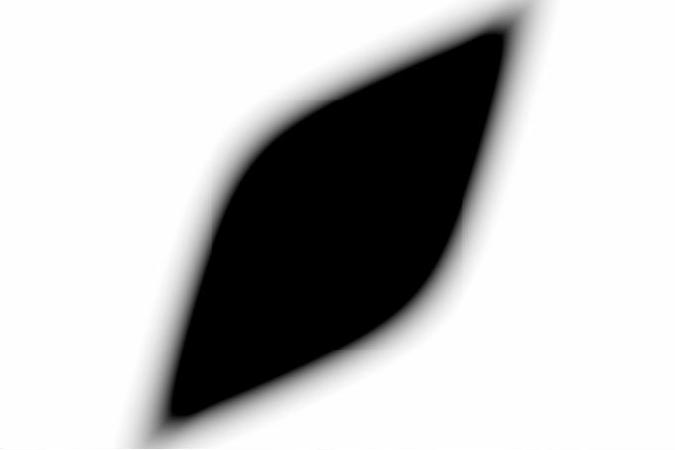}  \\
\includegraphics[width=0.45\textwidth]{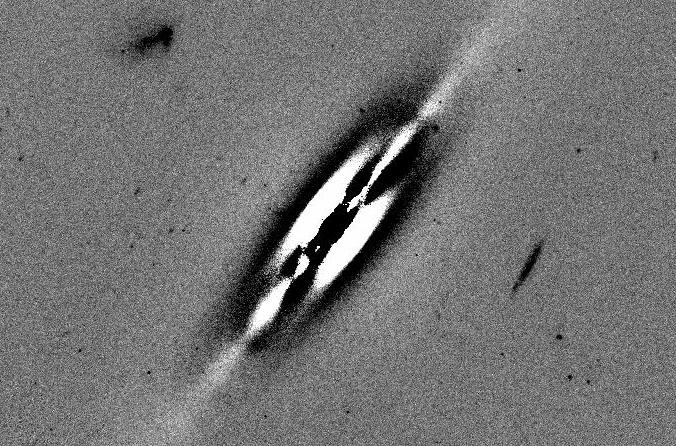} \\
\caption{Images in the $r$ band for SDSS J0827$+$2142. From the top: original image,
the central regions, best GALFIT models and residuals, as in Fig.~\ref{j0316img} in the
main text}
\label{j0827img}
\end{figure}

\begin{figure}
\centering
\hspace{-0.75cm}\includegraphics[width=0.43\textwidth]{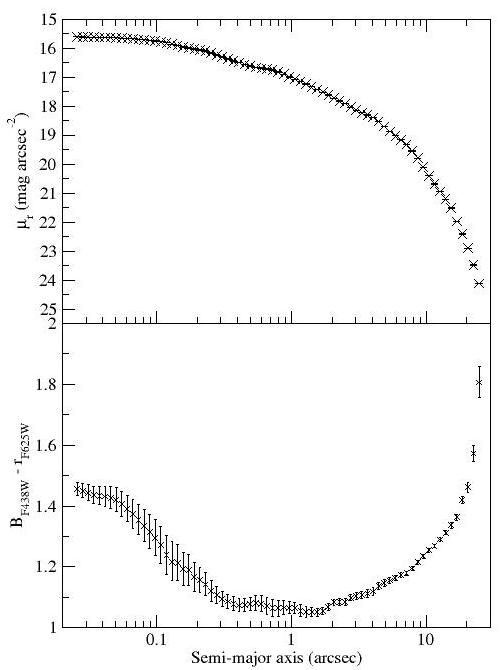}
\caption{Top: The $r$ band surface brightness profile (Vega magnitudes) for J0827+2142. Bottom:
The $B-r$ radial colour distribution}
\label{j0827prof}
\end{figure}

\begin{figure}
\center
\includegraphics[width=0.45\textwidth]{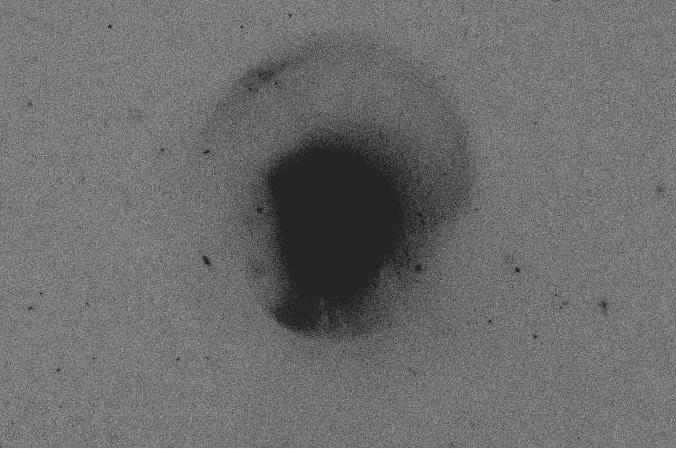}\\
\includegraphics[width=0.45\textwidth]{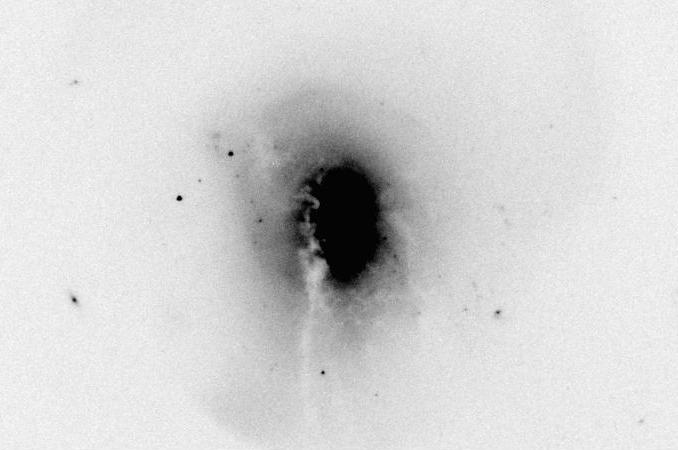}\\
\includegraphics[width=0.45\textwidth]{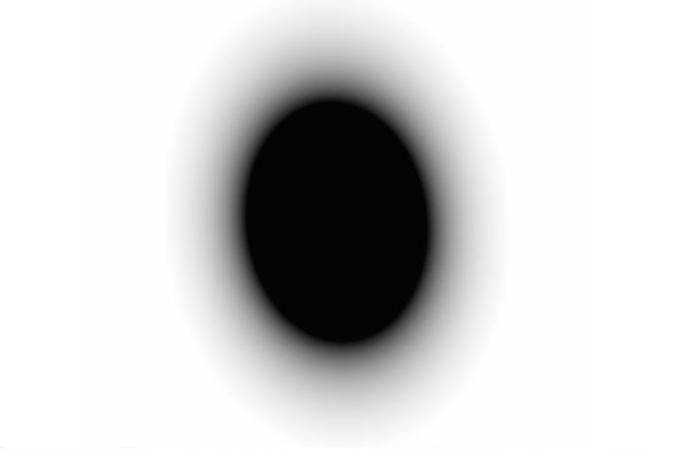} \\
\includegraphics[width=0.45\textwidth]{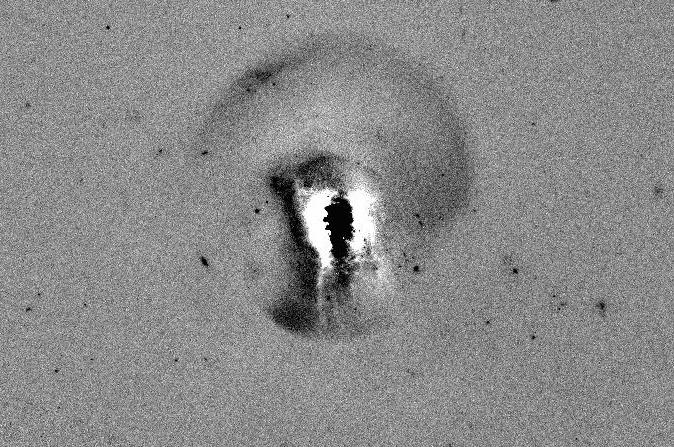}\\
\caption{Same as Fig.~\ref{j0827img} but for SDSS J0944+0429}
\label{j0944img}
\end{figure}

\begin{figure}
\centering
\hspace{-0.75cm}\includegraphics[width=0.43\textwidth]{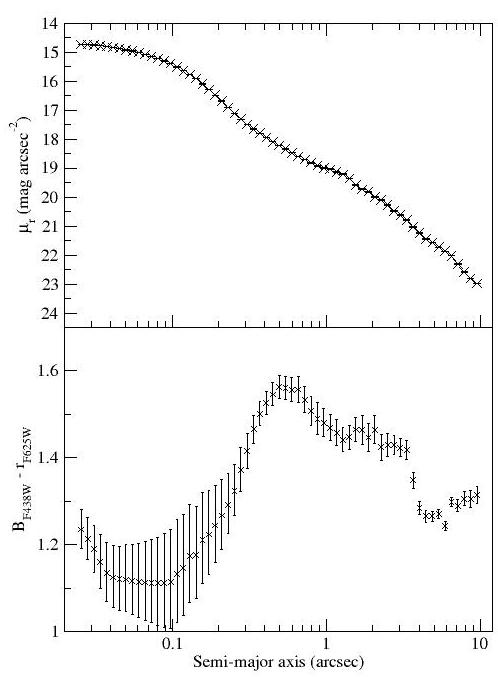}
\caption{Top: The $r$ band surface brightness profile (Vega magnitudes) for J0944+0429. Bottom:
The $B-r$ radial colour distribution}
\label{j0944prof}
\end{figure}

\begin{figure}
\center
\includegraphics[width=0.45\textwidth]{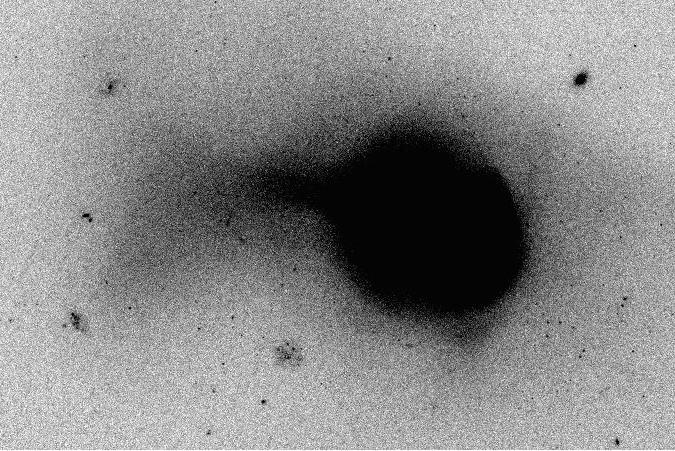}\\
\includegraphics[width=0.45\textwidth]{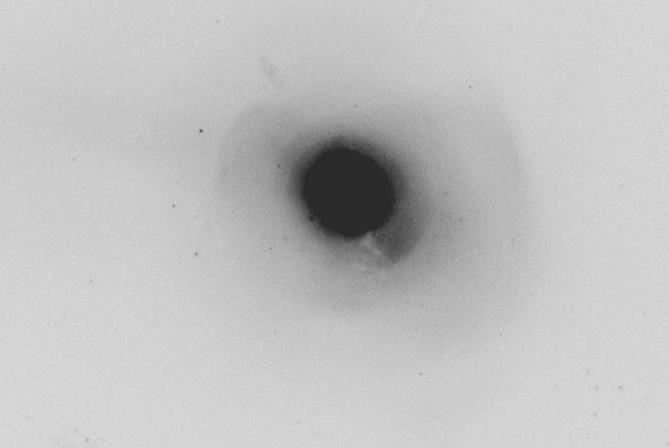}\\
\includegraphics[width=0.45\textwidth]{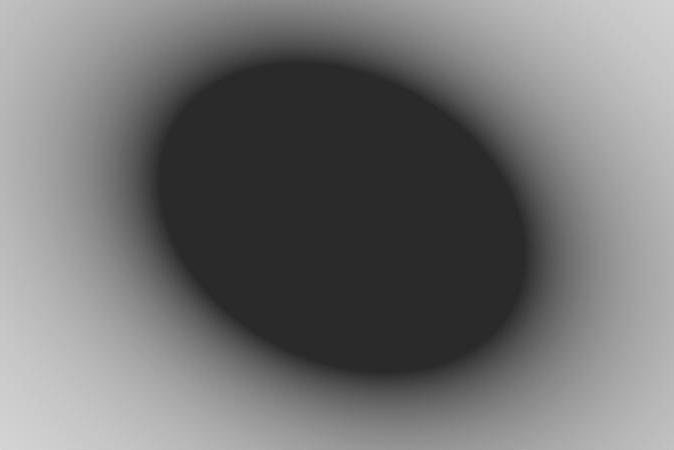} \\
\includegraphics[width=0.45\textwidth]{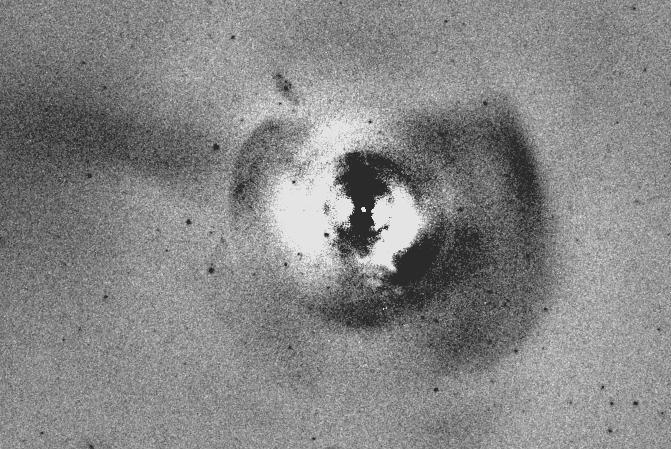}\\
\caption{Same as Fig.~\ref{j0827img} but for J1239+1226}
\label{j1239img}
\end{figure}

\begin{figure}
\centering
\hspace{-0.75cm}\includegraphics[width=0.43\textwidth]{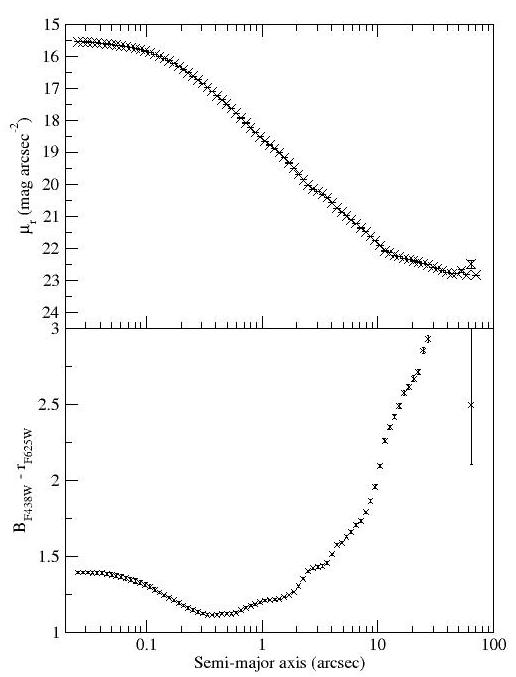}
\caption{Top: The $r$ band surface brightness profile (Vega magnitudes) for J1239+1226. Bottom:
The $B-r$ radial colour distribution}
\label{j1239prof}
\end{figure}

\begin{figure}
\center
\includegraphics[width=0.45\textwidth]{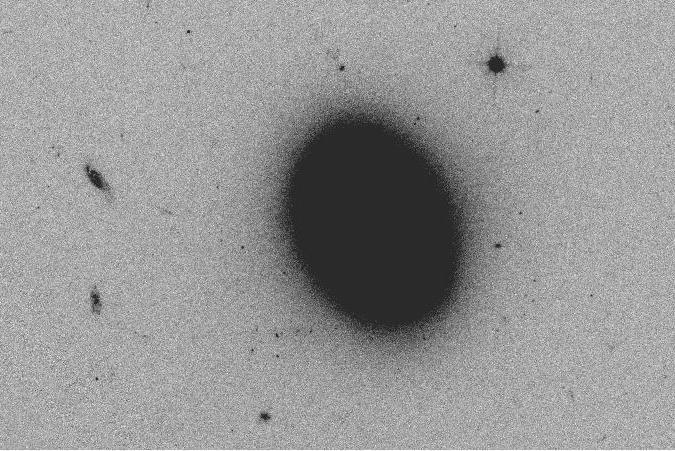}\\
\includegraphics[width=0.45\textwidth]{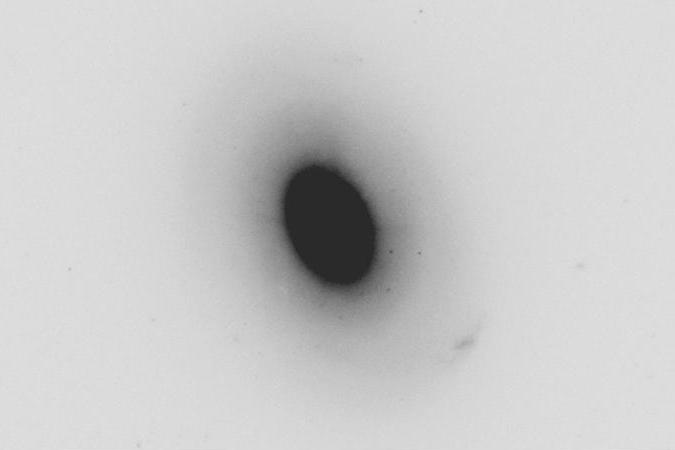}\\
\includegraphics[width=0.45\textwidth]{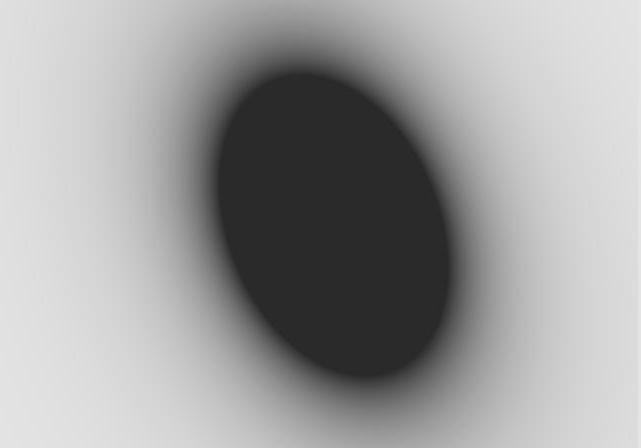} \\
\includegraphics[width=0.45\textwidth]{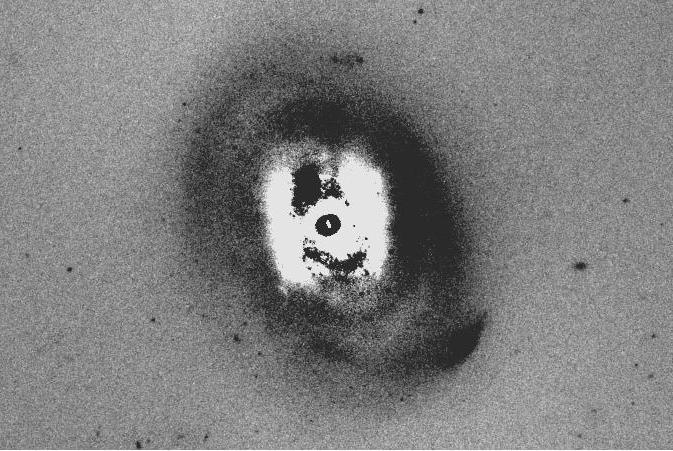}\\
\caption{As for Fig.~\ref{j0827img} but for SDSS J1305$+$5330}
\label{j1305img}
\end{figure}

\begin{figure}
\centering
\hspace{-0.75cm}\includegraphics[width=0.43\textwidth]{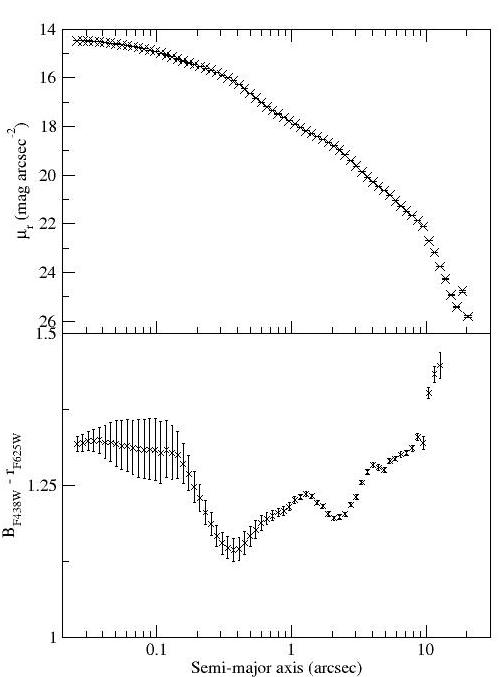}
\caption{Top: The $r$ band surface brightness profile (Vega magnitudes) for J1305+5330. Bottom:
The $B-r$ radial colour distribution}
\label{j1305prof}
\end{figure}

\begin{figure}
\center
\includegraphics[width=0.45\textwidth]{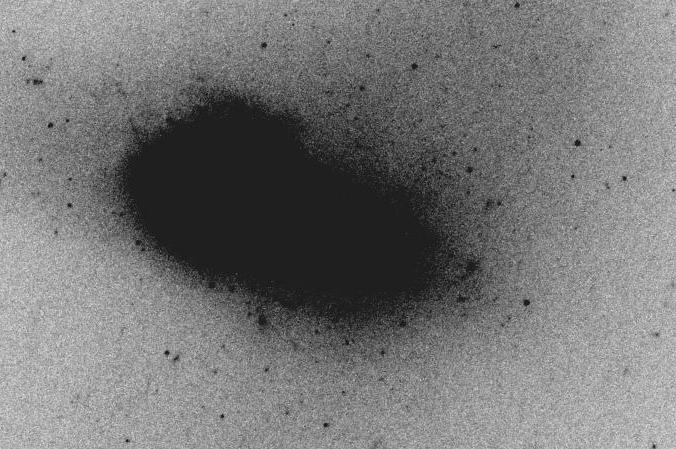}\\
\includegraphics[width=0.45\textwidth]{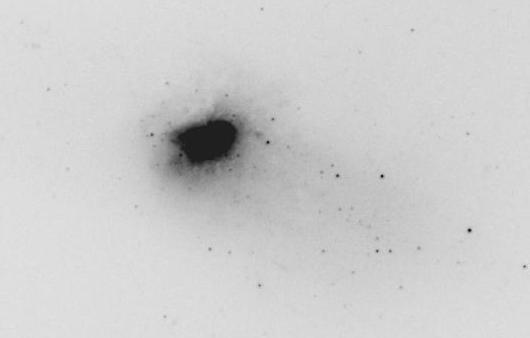}\\
\includegraphics[width=0.45\textwidth]{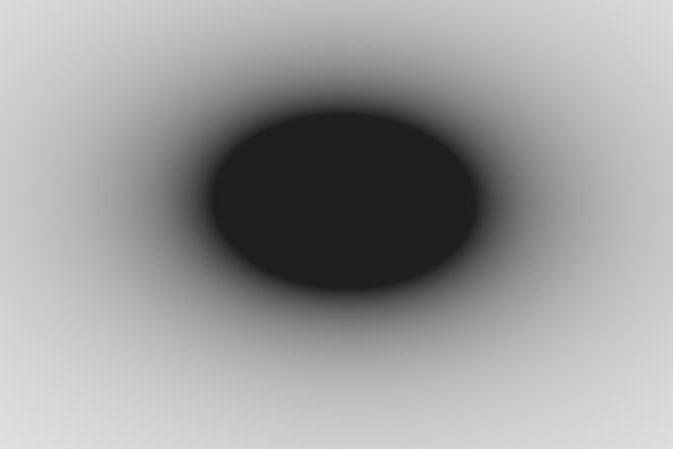} \\
\includegraphics[width=0.45\textwidth]{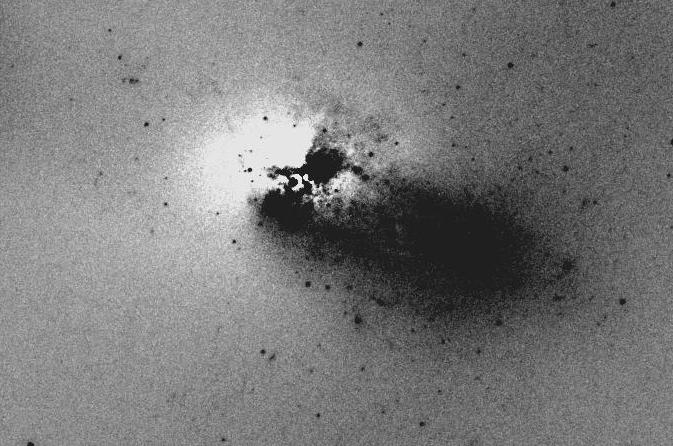}\\
\caption{Same as Fig.~\ref{j0827img} but for SDSS J1613$+$5103.}
\label{j1613img}
\end{figure}

\begin{figure}
\centering
\hspace{-0.75cm}\includegraphics[width=0.43\textwidth]{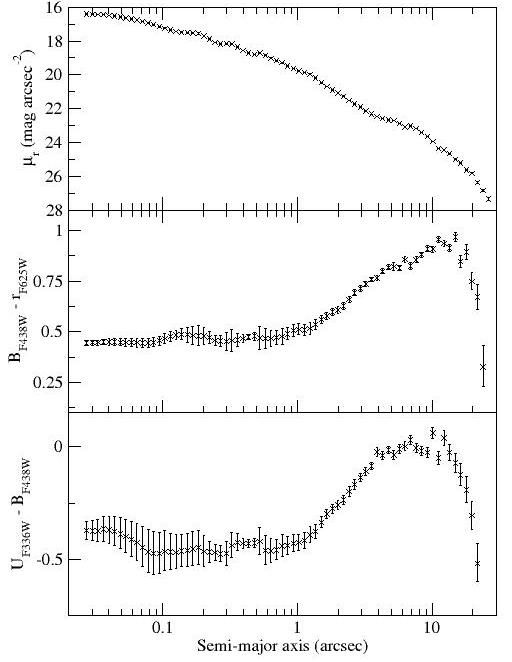}
\caption{Top: The $r$ band surface brightness profile (Vega magnitudes) for J1613+5103. Middle:
The $B-r$ radial colour distribution. Bottom: The $U-B$ colour profile.}
\label{j1613prof}
\end{figure}

\begin{figure}
\center
\includegraphics[width=0.45\textwidth]{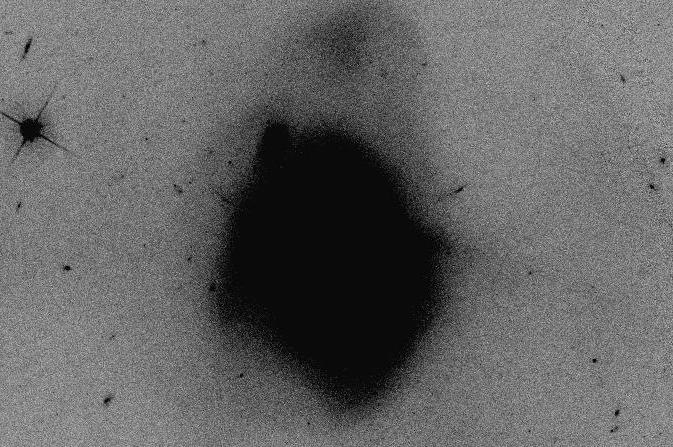}\\
\includegraphics[width=0.45\textwidth]{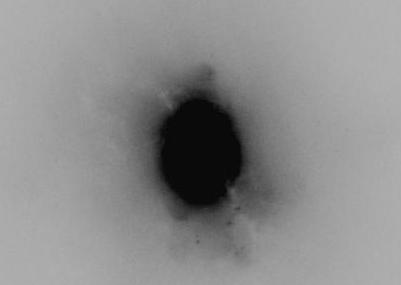}\\
\includegraphics[width=0.45\textwidth]{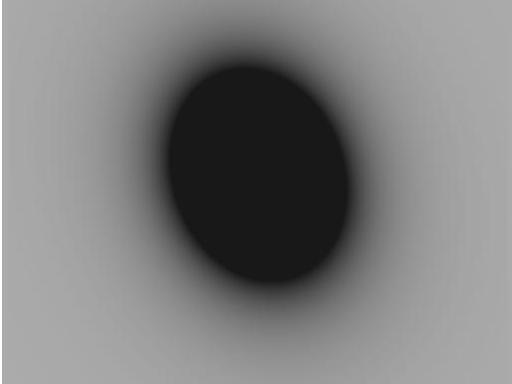} \\
\includegraphics[width=0.45\textwidth]{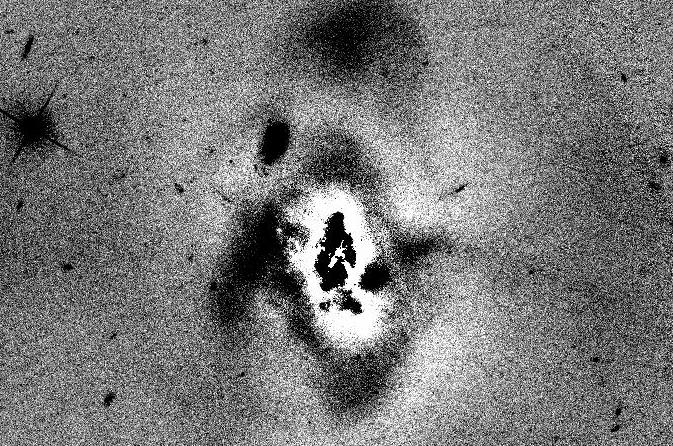}\\
\caption{As for Fig.~\ref{j0827img} but for  SDSS J1627$+$4328.}
\label{j1627img}
\end{figure}

\begin{figure}
\centering
\hspace{-0.75cm}\includegraphics[width=0.43\textwidth]{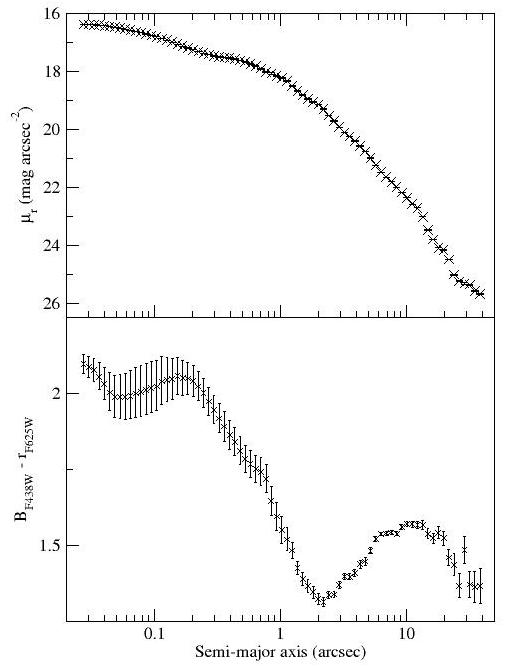}
\caption{Top: The $r$ band surface brightness profile (Vega magnitudes) for J1627+4328. Bottom:
The $B-r$ radial colour distribution}
\label{j1627prof}
\end{figure}

\begin{figure}
\center
\includegraphics[width=0.45\textwidth]{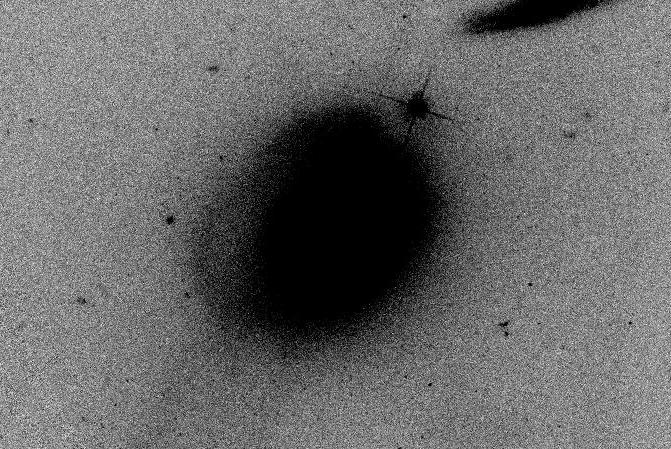}\\
\includegraphics[width=0.45\textwidth]{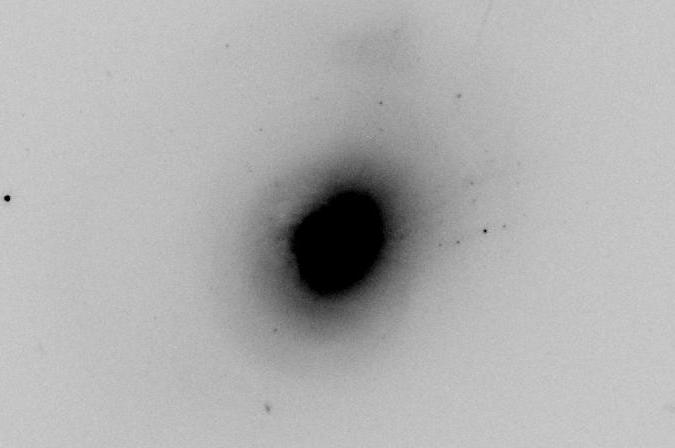}\\
\includegraphics[width=0.45\textwidth]{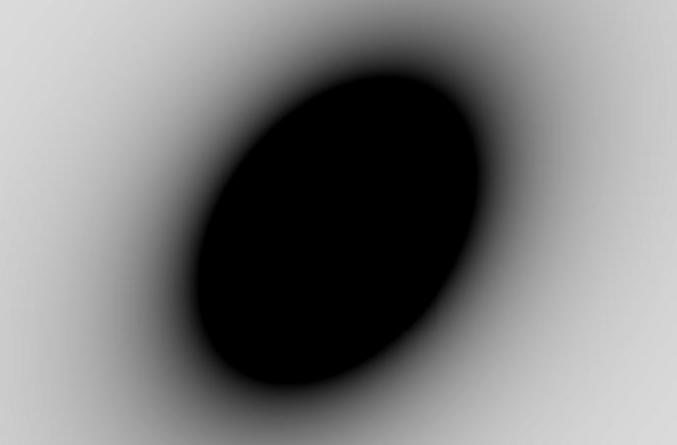} \\
\includegraphics[width=0.45\textwidth]{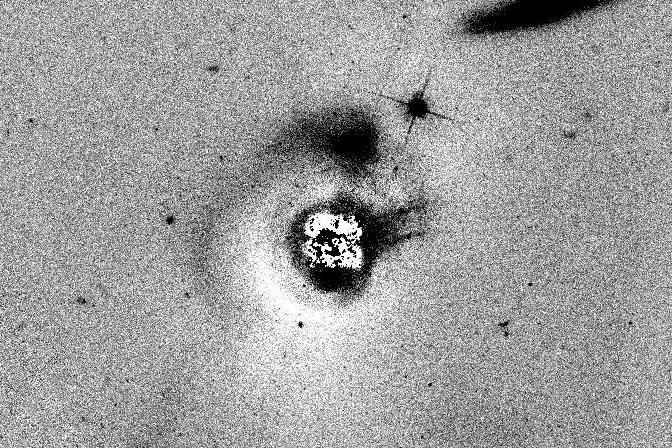}\\
\caption{As for Fig.~\ref{j0827img} but for SDSS J2255$+$0058.}
\label{j2255img}
\end{figure}

\begin{figure}
\centering
\hspace{-0.75cm}\includegraphics[width=0.43\textwidth]{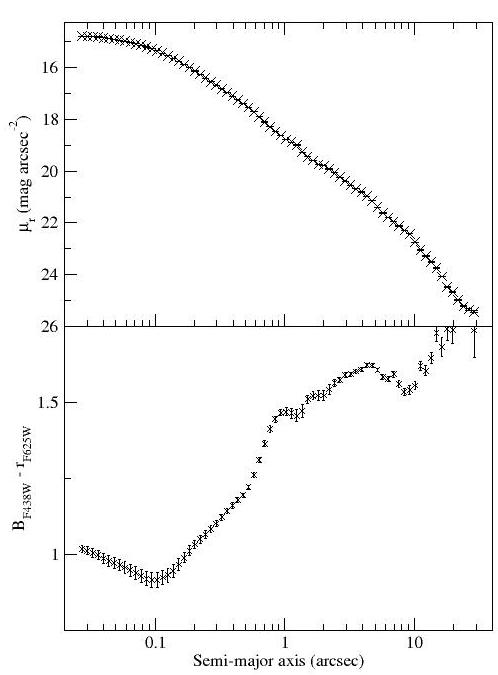}
\caption{Top: The $r$ band surface brightness profile (Vega magnitudes) for J2255+0058. Bottom:
The $B-r$ radial colour distribution}
\label{j2255prof}
\end{figure}

\clearpage

\section{Stellar Populations}
Here we plot the spectral energy distributions, {\tt STARLIGHT} models and the derived star formtion histories
for our galaxies.

\begin{figure}
\centering
\includegraphics[width=0.5\textwidth]{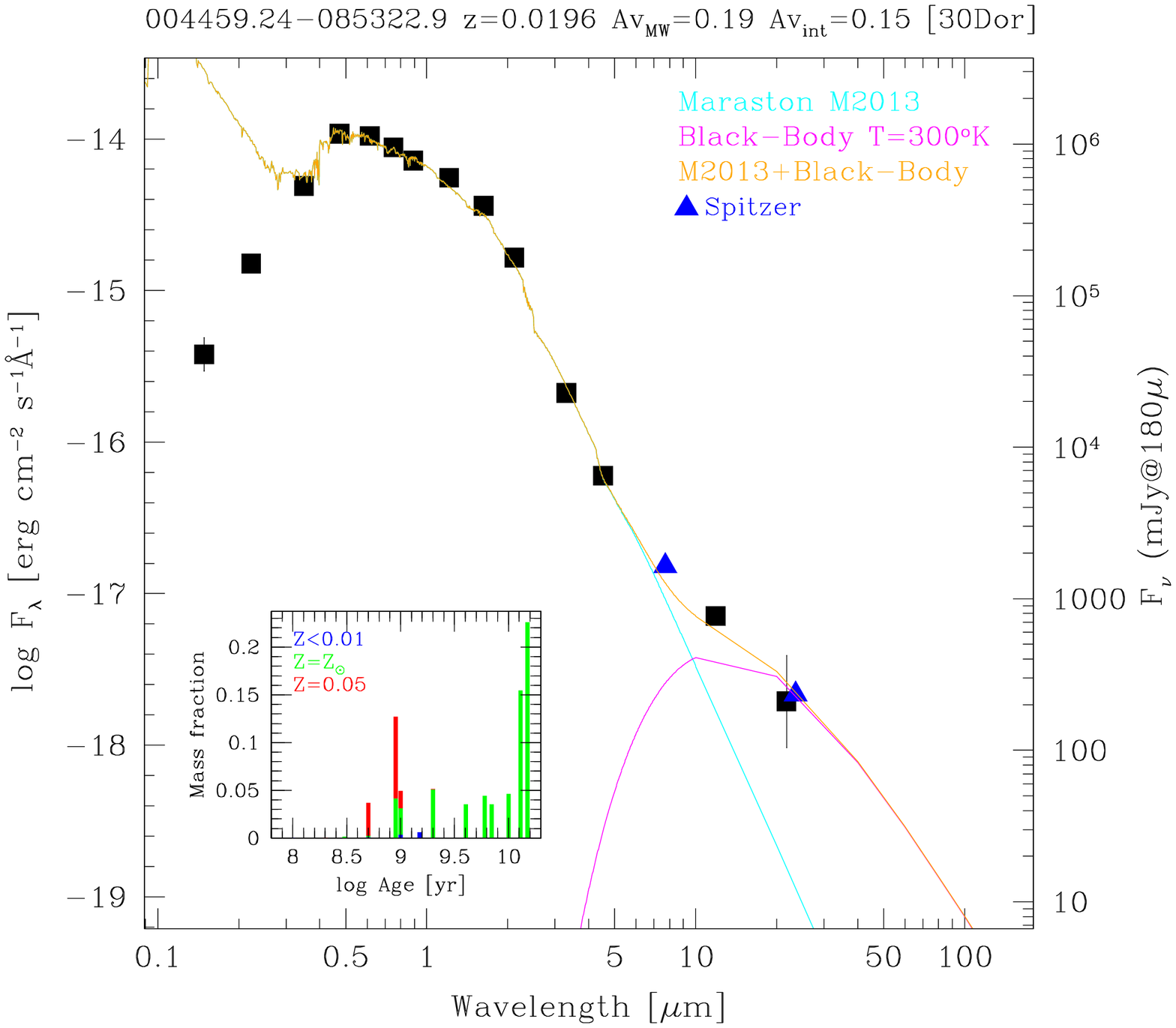}
\caption{Spectral energy distributions, SDSS spectra and models by Maraston (2005) for J0044--0853.
The inset shows the age, metallicity and fractional contributions of the stellar populations}
\label{j0044sed}
\end{figure}


\begin{figure}
\centering
\includegraphics[width=0.5\textwidth]{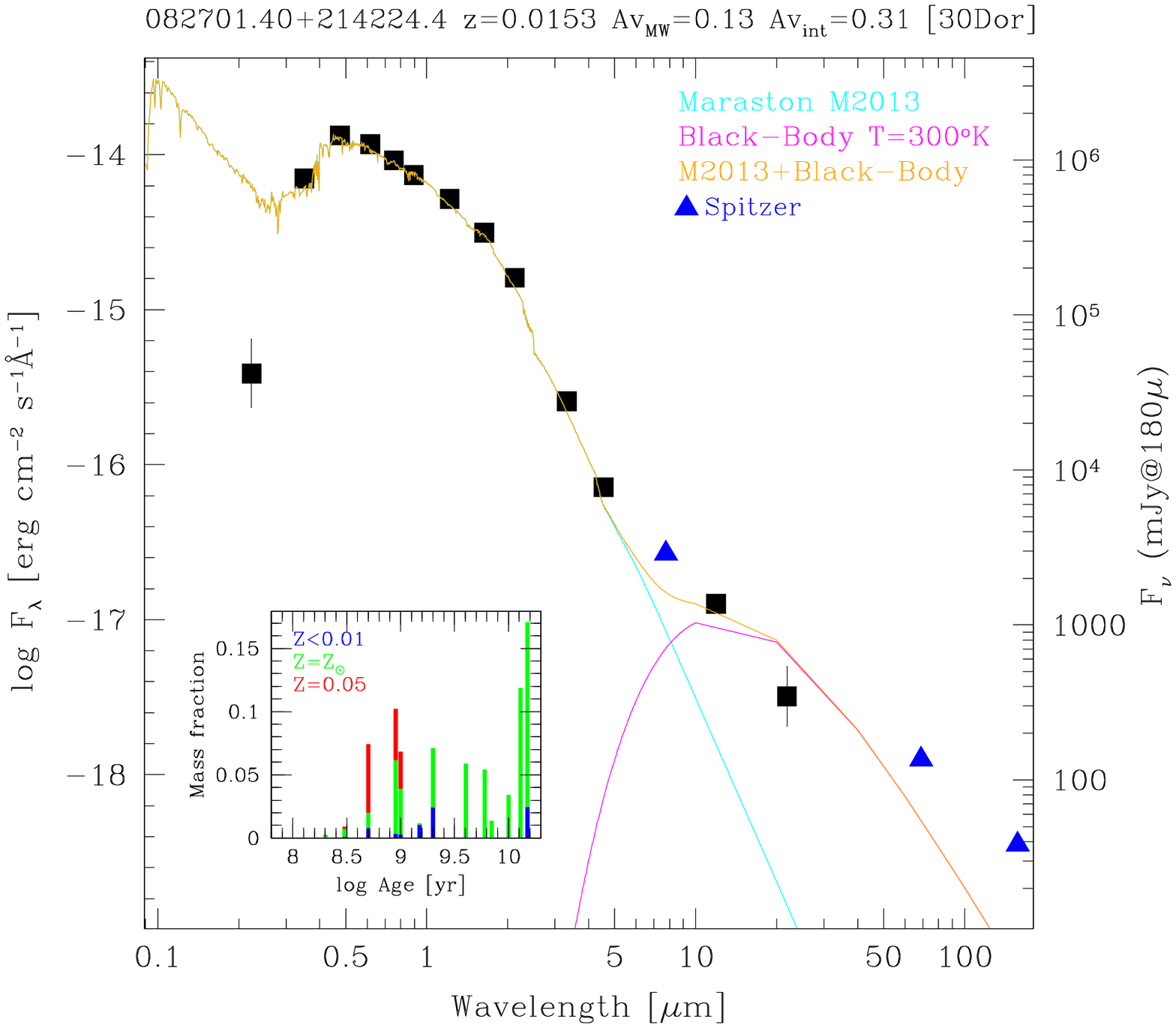}
\caption{Same as Fig.~\ref{j0044sed} but for J0827+2142}
\label{j0827sed}
\end{figure}

\begin{figure}
\centering
\includegraphics[width=0.5\textwidth]{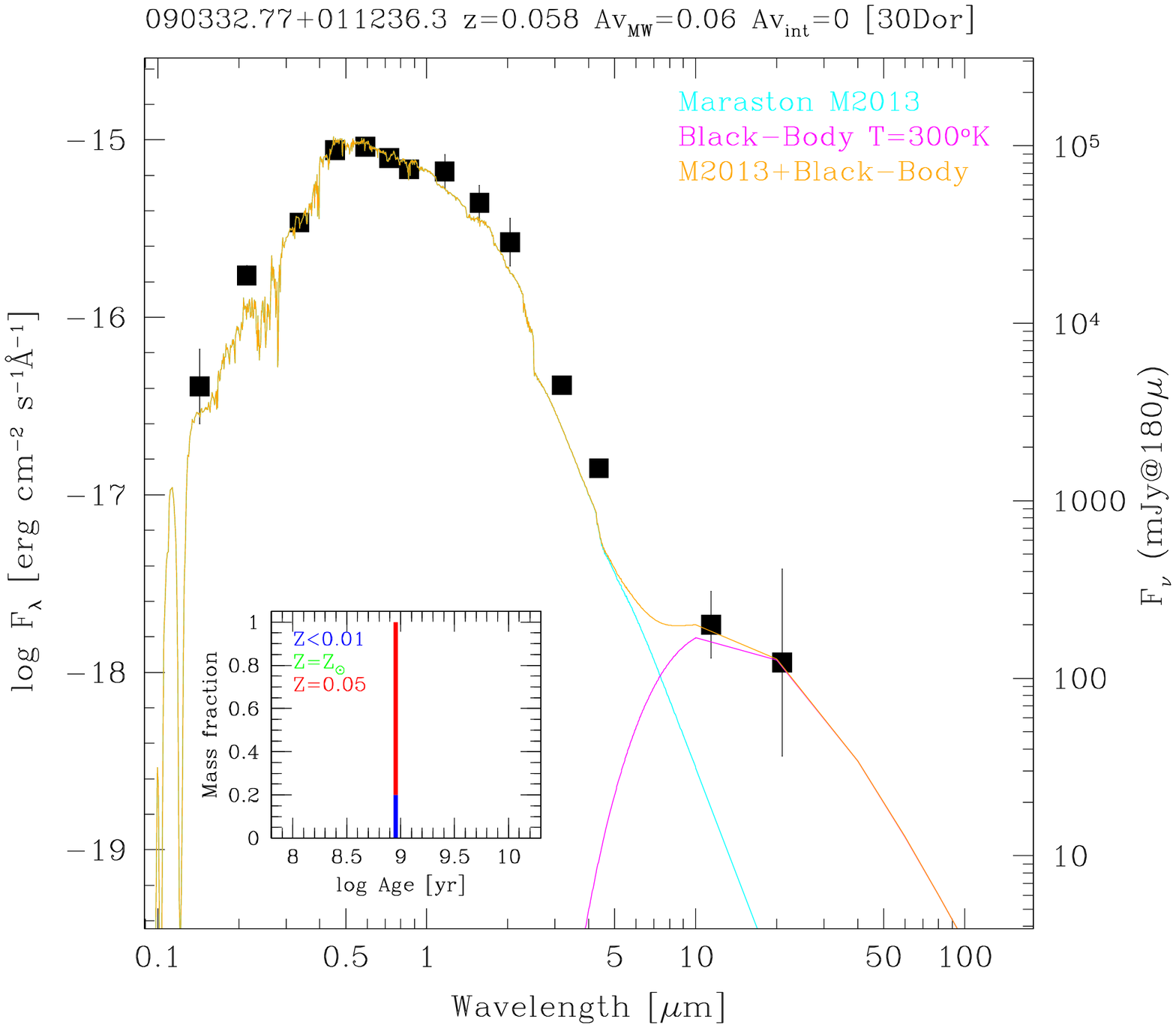}
\caption{Same as Fig.~\ref{j0044sed} but for J0903+0112}
\label{j0903sed}
\end{figure}

\begin{figure}
\centering
\includegraphics[width=0.5\textwidth]{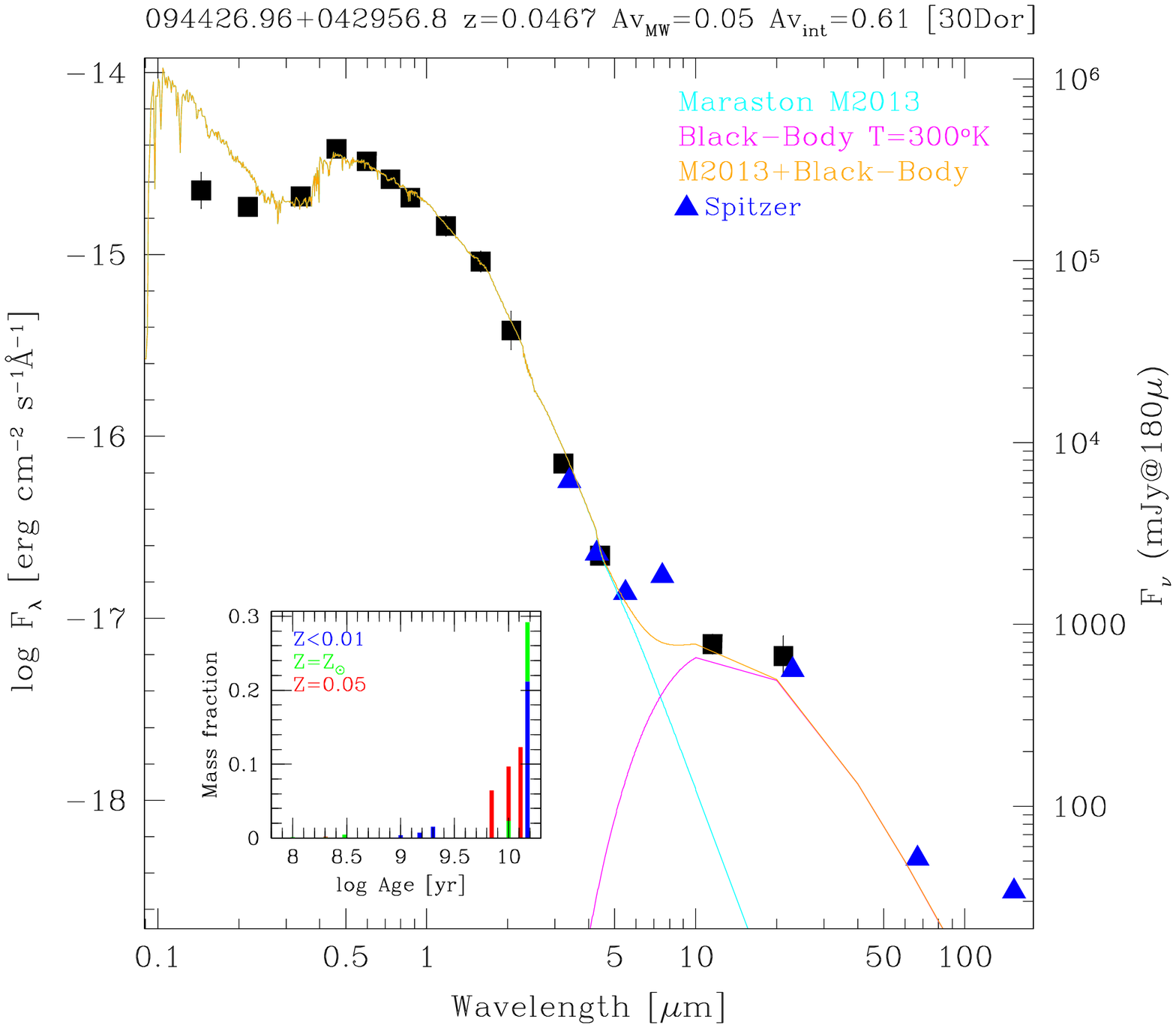}
\caption{Same as Fig.~\ref{j0044sed} but for J0944+0429}
\label{j0944sed}
\end{figure}

\begin{figure}
\centering
\includegraphics[width=0.5\textwidth]{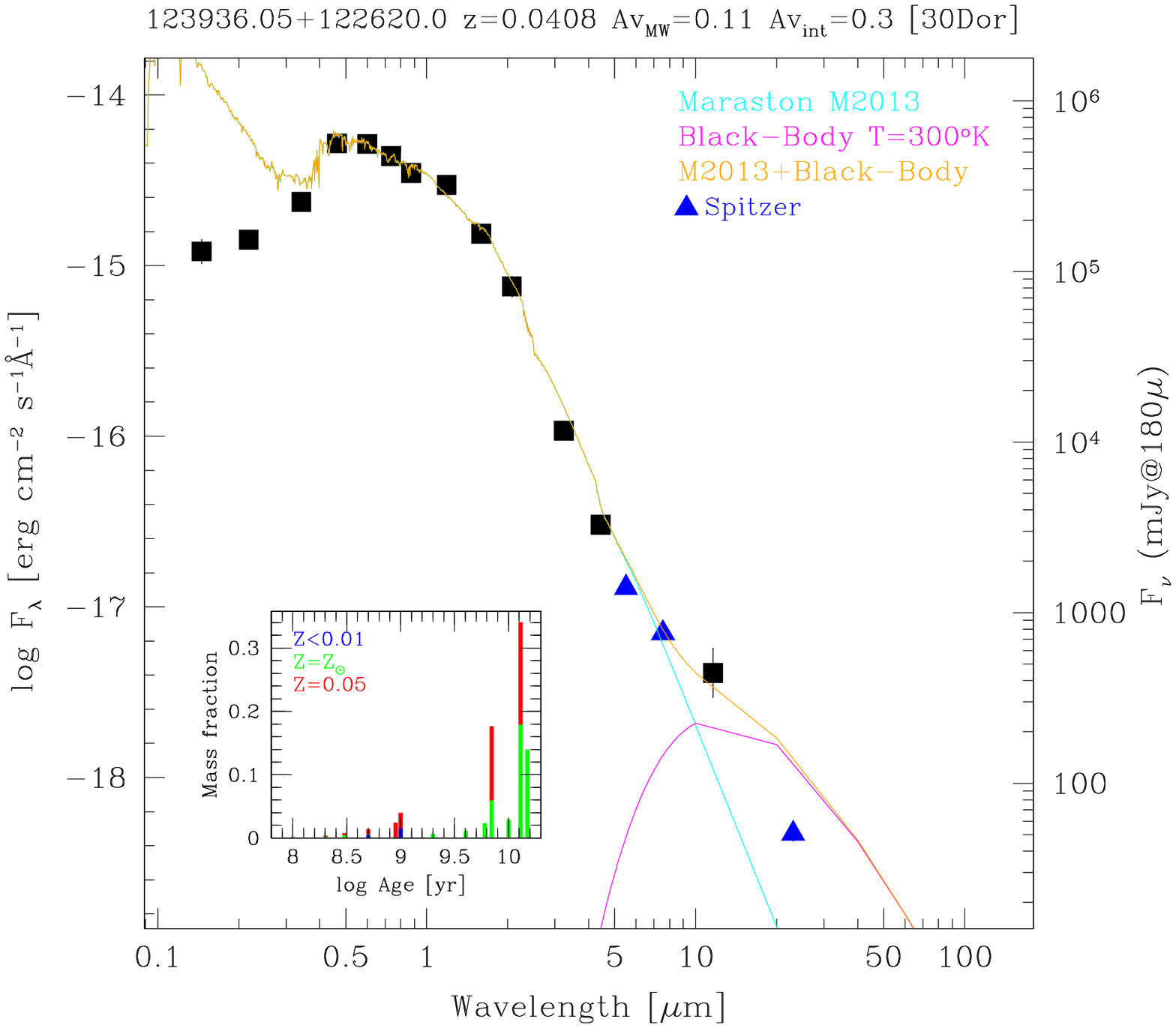}
\caption{Same as Fig~\ref{j0044sed} but for J1239+1226}
\label{j1239sed}
\end{figure}


\begin{figure}
\centering
\includegraphics[width=0.5\textwidth]{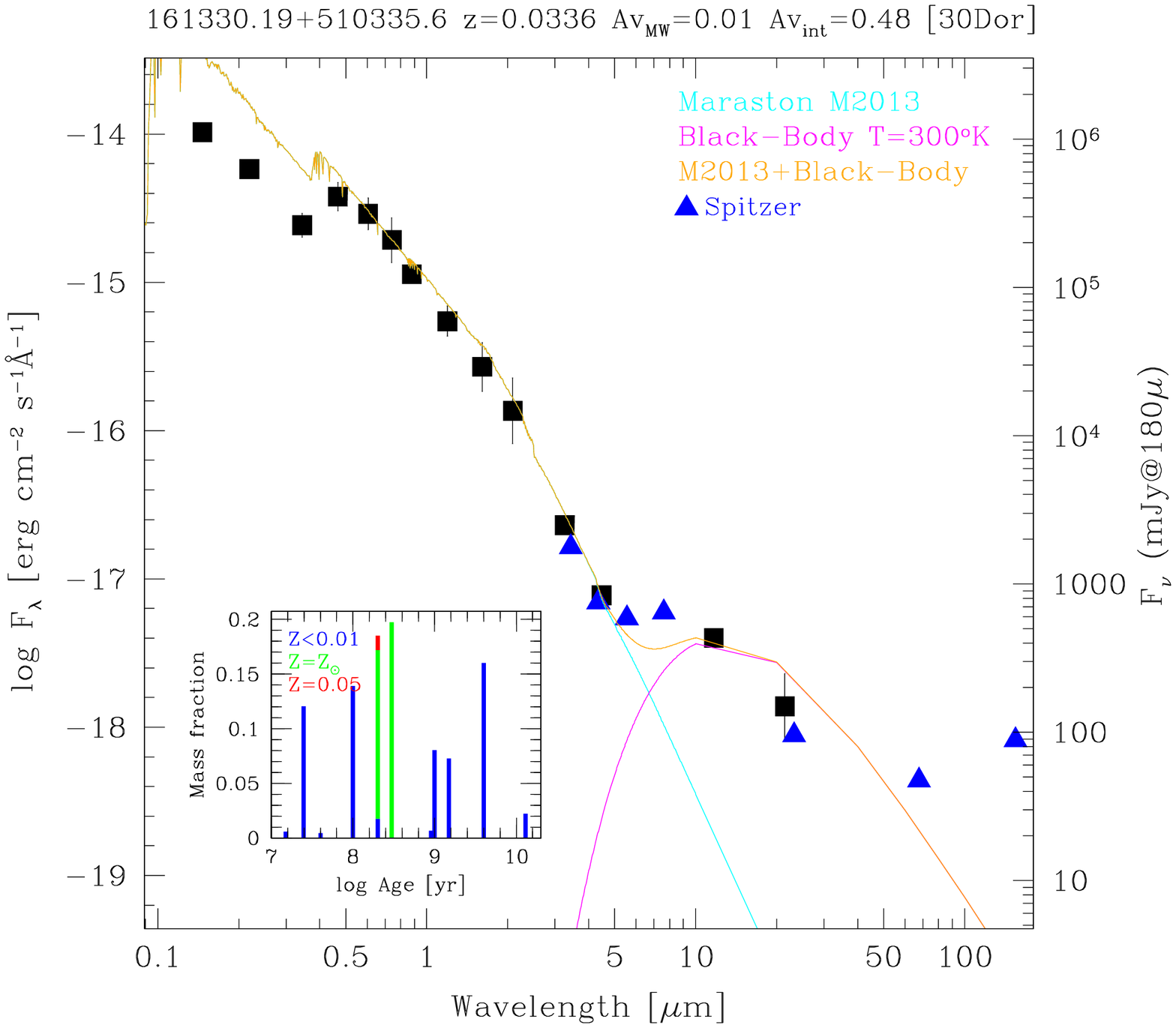}
\caption{Same as Fig.~\ref{j0044sed} but for J1613+5103}
\label{j1613sed}
\end{figure}

\begin{figure}
\centering
\includegraphics[width=0.5\textwidth]{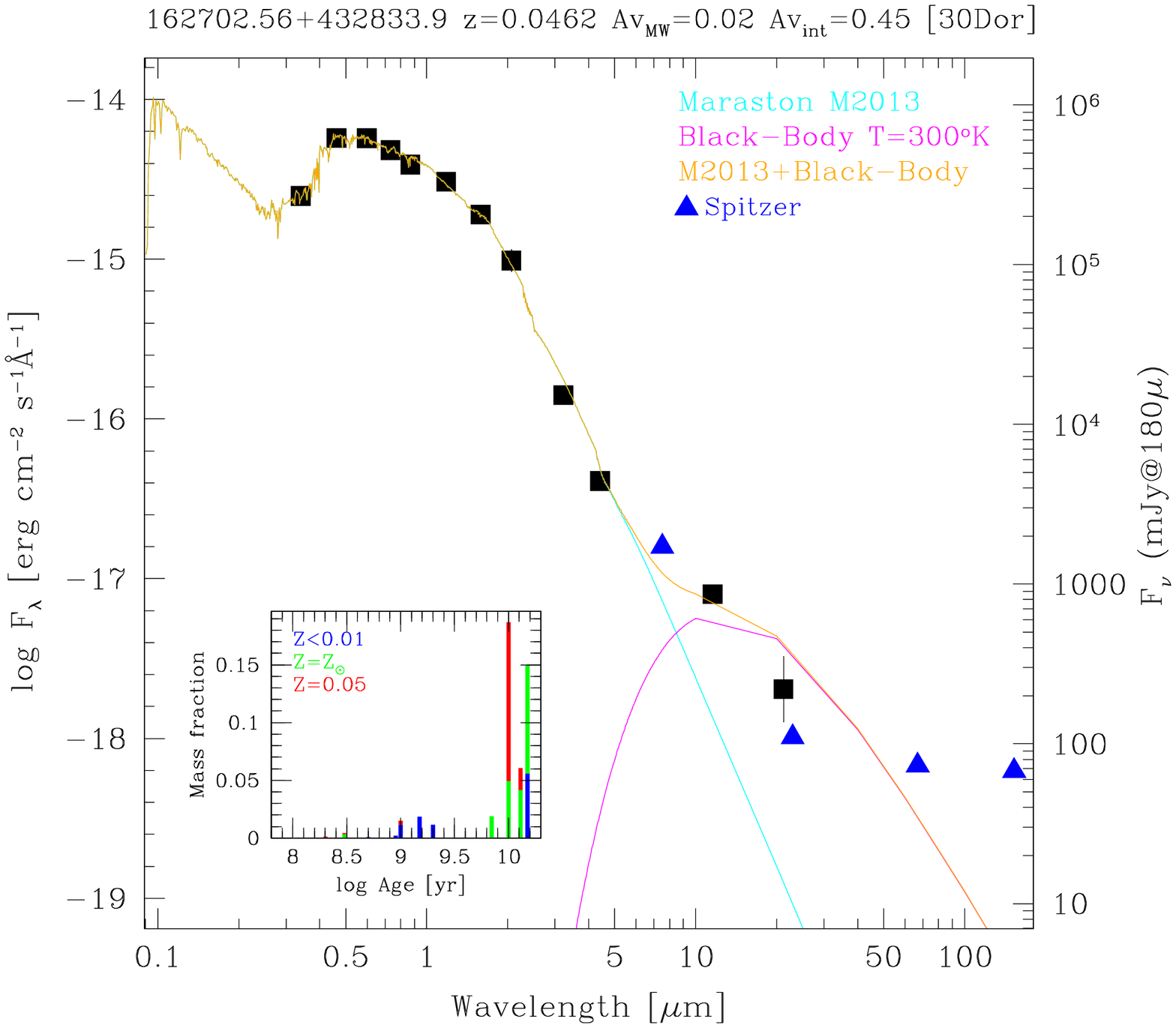}
\caption{Same as Fig.~\ref{j0044sed} but for J1627+4328}
\label{j1627sed}
\end{figure}

\begin{figure}
\centering
\includegraphics[width=0.5\textwidth]{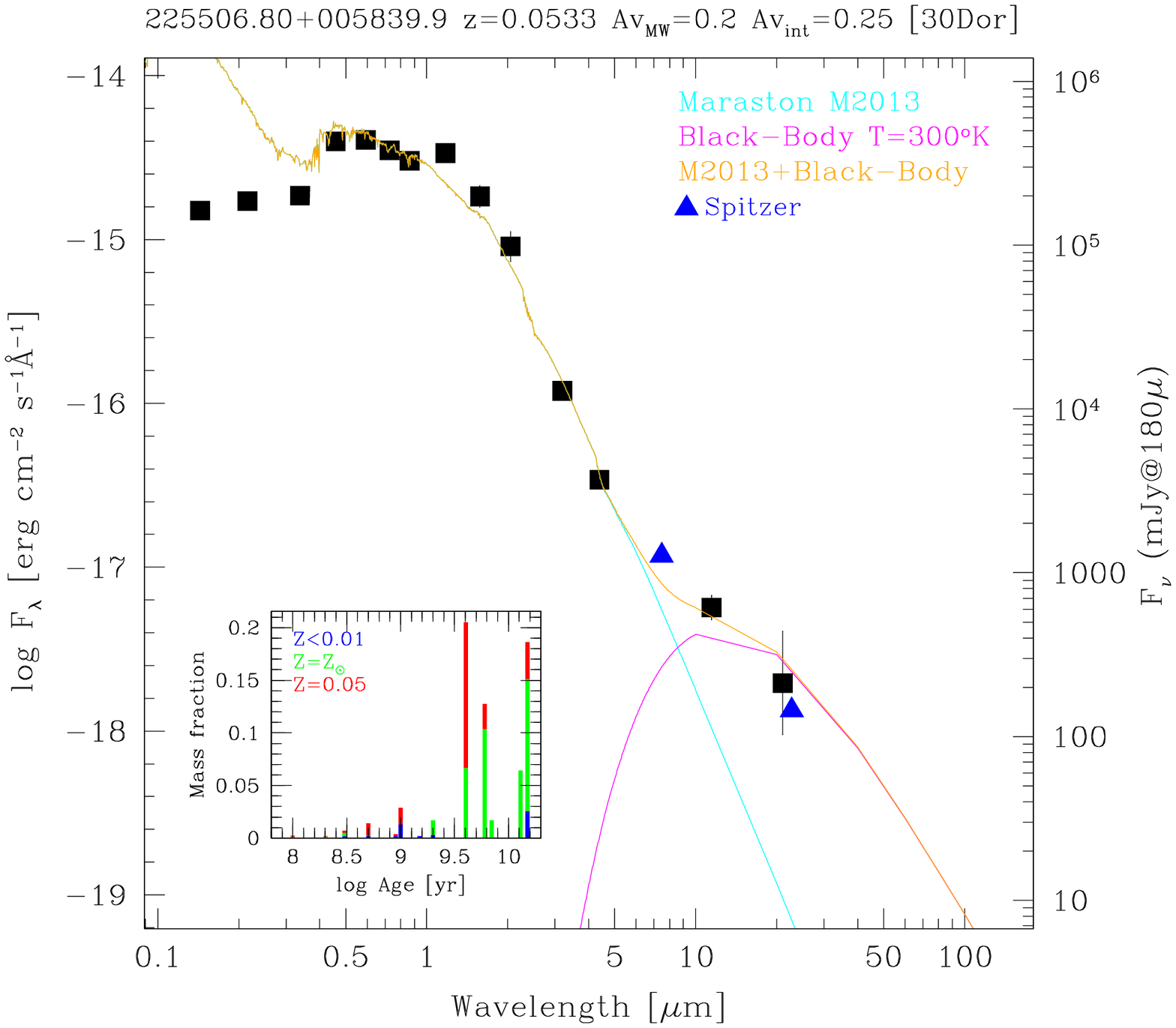}
\caption{Same as Fig.~\ref{j0044sed} but for J2255+0058}
\label{j2255sed}
\end{figure}

\end{document}